\documentclass[a4paper,aps,prd,10pt,preprintnumbers,showpacs,twocolumn,superscriptaddress,nofootinbib,amsmath,amssymb,floatfix]{revtex4-1}
\usepackage{graphicx}
\usepackage{xcolor}
\usepackage{cmap}
\usepackage[utf8]{inputenc}
\usepackage[T1]{fontenc}
\usepackage{hyperref}
\def\imo{i}

\def\K{{\cal K}}
\def\Order#1{{\cal O}\left(#1\right)}

\begin{document}
\title{Gravitational perturbations of a regular T-dulatiy inspired black hole: Quasinormal modes, excitation factors, and time-domain evolution}
\author{Bekir Can L{\"u}tf{\"u}o{\u{g}}lu}
\email{bekir.lutfuoglu@uhk.cz}
\affiliation{Department of Physics, Faculty of Science, University of Hradec Kr{\'a}lov{\'e}, Rokitansk{\'e}ho 62/26, 500 03 Hradec Kr{\'a}lov{\'e}, Czech Republic}

\author{Mardon~Abdullaev} 
\email{mardonabdullaev@gmail.com}
\affiliation{Institute of Fundamental and Applied Research, National Research University TIIAME, Kori Niyoziy 39, Tashkent 100000, Uzbekistan}
\affiliation{Tashkent State Technical University, Tashkent 100095, Uzbekistan}

\author{Javlon~Rayimbaev} 
\email{javlon@astrin.uz}
\affiliation{Institute of Theoretical Physics, National University of Uzbekistan, Tashkent 100174, Uzbekistan}
\affiliation{University of Tashkent for Applied Sciences, Gavhar Str. 1, Tashkent 700127, Uzbekistan}

\author{Shokhzod Jumaniyozov} 
\email{sh.jumaniyozov@newuu.uz}
\affiliation{Kimyo International University in Tashkent, Shota Rustaveli Street 156, Tashkent 100121, Uzbekistan}

\author{Shavkat~Karshiboev} \email{shavkatarshiboyev@samdpi.uz} \affiliation{Samarkand State Pedagogical Institute, Spitamen Shokh Street 166, Samarkand 140100, Uzbekistan} \affiliation{University of Economics and Pedagogy, 13 Islam Karimov St., Karshi, 180100, Uzbekistan}

\begin{abstract}
We study axial gravitational perturbations of the neutral regular black hole generated by a non-local, T-duality-inspired zero-point length and the associated gravitational self-energy.  In this geometry, the usual point source is replaced by a regular core, and the zero-point length controls the departure from the Schwarzschild limit.  We compute the fundamental quasinormal modes and several overtones with high-order WKB--Pad\'e methods, and we check the dominant mode with a direct time-domain evolution.  When the zero-point length is turned on, the real parts of the ADM-scaled frequencies increase for the gravitational modes with $\ell=2,3,4$, so the ringdown oscillates faster than in the Schwarzschild limit.  The damping rates change more mildly: they first grow slightly and then decrease near the largest deformation values considered here.  This behavior is consistent with the effective potential, whose barrier becomes higher as the deformation parameter increases.  We also compute the corresponding excitation factors and find that their magnitudes vary much less strongly than the quasinormal frequencies.
\end{abstract}
\maketitle
\section{Introduction}

The ringdown of a black hole is governed by its quasinormal spectrum, a set of damped oscillations fixed by the spacetime geometry and by the type of perturbation under consideration.  These modes provide one of the cleanest links between the theoretical structure of a compact object and the signal measured after a dynamical disturbance.  Their interpretation, accuracy and limitations have been developed in a large body of work~\cite{Berti:2009kk,Konoplya:2011qq,Nollert:1999ji,Kokkotas:1999bd,Bolokhov:2025rng}.  In particular, different parts of the spectrum probe different geometric regions: the near-horizon geometry can affect the first overtones~\cite{Konoplya:2022pbc}, while the eikonal modes are controlled by the properties of unstable null orbits~\cite{Cardoso:2008bp}.  For this reason, quasinormal modes are a sensitive diagnostic of regular, quantum-corrected or otherwise modified black-hole backgrounds.

Regular black holes are especially useful for this purpose because they keep the exterior causal structure of a black hole while replacing the central singularity with a finite core.  Such geometries arise in several effective settings, including nonlinear electrodynamics, asymptotic-safety inspired models, noncommutative constructions, vacuum polarization effects and phenomenological regularizations of the gravitational source~\cite{Ayon-Beato:1998hmi,Bronnikov:2000vy,Bonanno:2000ep,Konoplya:2024kih,Hayward:2005gi,Dymnikova:1992ux,Nicolini:2005vd,Bronnikov:2024izh,Konoplya:2025ect,Spina:2025wxb,Bonanno:2025dry,Bolokhov:2024sdy}.  Their spectra, scattering properties and evaporation channels show that even when the horizon remains close to the Schwarzschild one, the deformation of the effective potential outside the horizon can leave a noticeable imprint on the ringing frequencies, damping rates and transmission probabilities~\cite{Macedo:2016yyo,MahdavianYekta:2019pol,Konoplya:2023aph,Guo:2024jhg,Li:2014fka,Skvortsova:2024wly,Pedraza:2021hzw,Bolokhov:2025fto,Bolokhov:2023ruj,Skvortsova:2026unq,Pedrotti:2024znu,Vagnozzi:2022moj,Calza:2024xdh,Konoplya:2023ppx,Flachi:2012nv,Arbey:2021jif,Bolokhov:2026eqf,Held:2019xde,Gingrich:2024tuf,Panotopoulos:2019qjk,Dubinsky:2026wcv,Skvortsova:2025cah,Huang:2023aet,Yang:2021cvh,Arbey:2021yke,Lin:2013ofa,Lopez:2022uie,Arbey:2026koc}.

The background studied here is the neutral regular black hole proposed by Jusufi and Singleton~\cite{Jusufi:2025selfenergy}.  Its regularization is produced by the source endowed with a zero-point length $l_0$ motivated by a non-local, T-duality-inspired description, and the same scale regularizes the gravitational self-energy of the configuration.  The resulting self-energy contribution modifies the effective mass profile.  This makes the solution an interesting laboratory for black-hole spectroscopy, because the parameter $l_0$ simultaneously controls the size of the nonsingular core and the departure of the exterior potential from its Schwarzschild form.  Grey-body factors and Hawking evaporation for this spacetime were considered in Ref.~\cite{3164933}, while quasinormal modes and excitation factors of test scalar, electromagnetic and massless Dirac test fields were analyzed in~\cite{Skvortsova:2026testfields}.

The aim of the present paper is to complete this picture by studying gravitational perturbations of the same self-energy regular black hole.  This step is necessary because gravitational modes describe oscillations of the geometry itself, rather than the response of external test fields, and are therefore the sector most directly related to black-hole ringdown observations.  We compute the quasinormal frequencies for several multipoles and overtones using high-order WKB expansions with Pad\'e resummation, and we also evaluate the corresponding excitation factors.  The latter are the source-independent residues of the frequency-domain Green function at the quasinormal poles, and therefore provide information complementary to the frequencies: they quantify the intrinsic strength of each mode in a fixed scattering normalization.  By comparing these data with the test-field results, one can separate features that are generic consequences of the self-energy deformation from those that are specific to the gravitational sector.

The paper is organized as follows.  In Sec.~\ref{sec:wavelike}, we review the self-energy regular black-hole geometry, its ADM mass, and the axial gravitational effective potential.  Section~\ref{sec:methods} summarizes the WKB--Pad\'e and time-domain methods used to extract the spectra.  In Sec.~\ref{sec:qnm} we present the fundamental and overtone quasinormal modes, discuss their dependence on $l_0/M_{\rm ADM}$, and compare the WKB results with the time-domain evolution and late-time tail.  Section~\ref{sec:excitation} is devoted to the excitation factors and to the near-Schwarzschild check of the fundamental residues and frequencies.  Finally, Sec.~\ref{sec:conclusions} summarizes the main results and outlines possible observational and scattering extensions.

\section{Black hole spacetime}\label{sec:wavelike}

The spacetime studied below is the neutral asymptotically flat regular black hole constructed by Jusufi and Singleton from a regularized gravitational self-energy~\cite{Jusufi:2025selfenergy}.  The starting point is a non-local, T-duality-inspired description in which the usual point source is replaced by an effective distribution with a zero-point length $l_0$.  At the Newtonian level, this prescription replaces the singular potential by
\begin{equation}
 \Phi(r)=-\frac{M}{\sqrt{r^2+l_0^2}},
\end{equation}
The origin of this replacement can be seen from the proper-time duality used to implement a minimum length.  The corresponding momentum-space propagator is modified to
\begin{equation}\label{dual-propagator}
 \widetilde G(k)=-\frac{l_0}{\sqrt{k^2+m_0^2}}
 K_1\!\left(l_0\sqrt{k^2+m_0^2}\right),
\end{equation}
where $K_1$ is a modified Bessel function.  For a massless mediator, Eq.~(\ref{dual-propagator}) approaches the ordinary $-1/k^2$ propagator at small momentum, whereas its large-momentum behavior is exponentially suppressed.  Its static Fourier transform therefore replaces the point-source potential $-M/r$ by the finite expression above.  This is the sense in which the zero-point length produces a spatially smeared, non-local source rather than an additional material charge~\cite{Padmanabhan:1997zpl,Nicolini:2022zpl,Nicolini:2019tduality,Gaete:2022tduality,JusufiNicolini:2025geodesic,Jusufi:2022three,Jusufi:2023collapse}. So that Poisson's equation gives the regular bare density
\begin{eqnarray}\label{bare-density}
 \rho_{\rm bare}(r)&=&\frac{3Ml_0^2}{4\pi\left(r^2+l_0^2\right)^{5/2}},\nonumber\\
 4\pi\int_0^\infty \rho_{\rm bare}(r)r^2dr&=&M .
\end{eqnarray}
Explicitly, $\rho_{\rm bare}=(4\pi r^2)^{-1}d[r^2\Phi'(r)]/dr$; hence the first density profile is fixed directly by the regularized potential, while the second line verifies that its integrated bare mass is $M$.

The underlying non-local model can be summarized by an Einstein-Hilbert term coupled to a bare matter Lagrangian and to an additional gravitational-self-energy contribution,
\begin{equation}
 S=S_{\rm EH}+S_{\rm bare}+S_{\rm GSE},
 \qquad
 S_{\rm EH}=\frac{1}{16\pi}\int d^4x\sqrt{-g}\,R .
\end{equation}
More explicitly, the self-energy sector is represented schematically by
\begin{eqnarray}\label{nonlocal-kernel}
 S_{\rm GSE}&=&\frac{1}{16\pi}\int d^4x\sqrt{-g(x)}\,
 \mathcal L_{\rm GSE}(x),\nonumber\\
 \mathcal L_{\rm GSE}(x)&=&\int d^4y\sqrt{-g(y)}\, \mathcal K(x-y)\mathcal L_{\rm bare}(y),\nonumber\\
 \mathcal K(x-y)&=&\delta(x^0-y^0)\,\mathcal R(\boldsymbol z),
 \qquad \boldsymbol z=\boldsymbol x-\boldsymbol y,\nonumber\\
 \mathcal R(\boldsymbol z)&\propto&
 \frac{1}{\sqrt{|\boldsymbol z|^2+l_0^2}} .
\end{eqnarray}
Thus the kernel is local in time for the static problem but non-local in space: the source at $\boldsymbol y$ contributes to the effective density at $\boldsymbol x$ over a region controlled by $l_0$.  This construction follows the general idea that non-local gravity can represent gravitational-field contributions through an integral transform of the matter source~\cite{Hehl:2009nonlocal,JusufiSingletonLobo:2026collapse}.  In the present model the explicit self-energy density is obtained in the Newtonian limit and is then used as an effective source in the strong-field geometry; a first-principles strong-field non-local Lagrangian is not assumed~\cite{Jusufi:2025selfenergy}.

In the static spherical limit, the regularized gravitational field has magnitude $|\boldsymbol g|=|\Phi'(r)|=Mr/(r^2+l_0^2)^{3/2}$.  Identifying the effective field-energy density with the finite scalar $|\boldsymbol g|^2/(8\pi)$ gives
\begin{equation}\label{gse-density}
 \rho_{\rm GSE}(r)=\frac{M^2r^2}{8\pi\left(r^2+l_0^2\right)^3} .
\end{equation}
This explains why the two profiles have different origins: $\rho_{\rm bare}$ follows from Poisson's equation for the smeared material source, whereas $\rho_{\rm GSE}$ represents the regularized energy stored in its gravitational field~\cite{Jusufi:2025selfenergy,Nicolini:2019tduality}. Thus, the effective source entering Einstein's equations is an anisotropic fluid,
\begin{equation}
 T^{\mu}_{\ \nu}={\rm diag}[-\rho,p_r,p_t,p_t],
 \qquad
 \rho=\rho_{\rm bare}+\rho_{\rm GSE},
\end{equation}
with radial equation of state $p_r=-\rho$ and tangential pressure fixed by stress-energy conservation,
\begin{equation}
 p_t=-\rho-\frac{r}{2}\frac{d\rho}{dr} .
\end{equation}
These pressure relations are fixed by the geometry rather than chosen independently.  For the Schwarzschild-like gauge $g_{tt}=-1/g_{rr}=-f(r)$, the Einstein tensor obeys $G^t_{\ t}=G^r_{\ r}$, which implies $T^t_{\ t}=T^r_{\ r}$ and hence $p_r=-\rho$.  The remaining conservation equation $\nabla_\mu T^\mu_{\ r}=0$ then yields the stated expression for $p_t$.  The radial dependence of $p_t$ is therefore the effective anisotropic response required to support the two regular density profiles. The density is positive and finite, while the anisotropic pressures encode the effective non-local gravitational contribution; as in regular black-hole models, the standard pointwise energy conditions can be violated in the inner core.

The line element can be written in Schwarzschild-like coordinates as
\begin{eqnarray}\label{metric}
 ds^2&=&-f(r)dt^2+\frac{dr^2}{f(r)}+r^2(d\theta^2+\sin^2\theta d\phi^2),\nonumber\\
 f(r)&=&1-\frac{2m(r)}{r} .
\end{eqnarray}
The mass function is obtained by integrating the total density, because the $t$--$t$ Einstein equation reduces to $m'(r)=4\pi r^2\rho(r)$.  Regularity fixes $m(0)=0$, so integrating $\rho_{\rm bare}+\rho_{\rm GSE}$ gives
\begin{widetext}
\begin{equation}\label{mass-function}
 m(r)=4\pi\int_0^r \rho(x)x^2dx
 =\frac{Mr^3}{\left(r^2+l_0^2\right)^{3/2}}
 +\frac{3M^2}{16l_0}\tan^{-1}\left(\frac{r}{l_0}\right)
 -\frac{3M^2l_0^2r}{16\left(r^2+l_0^2\right)^2}
 -\frac{5M^2r^3}{16\left(r^2+l_0^2\right)^2} .
\end{equation}
\end{widetext}
Consequently, the metric function used in the numerical analysis is
\begin{widetext}
\begin{eqnarray}\label{metric-function}
 f(r)&=&1-\frac{2 M r^2}{\left(l_0^2+r^2\right)^{3/2}}
 -\frac{3M^2}{8l_0 r}\tan^{-1}\left(\frac{r}{l_0}\right)
 +\frac{3l_0^2M^2}{8\left(l_0^2+r^2\right)^2}
 +\frac{5M^2r^2}{8\left(l_0^2+r^2\right)^2} .
\end{eqnarray}
\end{widetext}
Here $M$ is the bare mass scale and $l_0$ is the zero-point length.  The ADM mass is the asymptotic value of Eq.~(\ref{mass-function}),
\begin{equation}\label{ADMmass-section}
 M_{\rm ADM}=M+\frac{3\pi M^2}{32l_0},
\end{equation}
and the large-distance expansion is
\begin{equation}\label{large-r-f}
 f(r)=1-\frac{2M_{\rm ADM}}{r}+\frac{M^2}{r^2}
 +\frac{3Ml_0^2}{r^3}-\frac{M^2l_0^2}{r^4}+\Order{r^{-5}} .
\end{equation}
Near the origin, the metric is regular and has a de-Sitter-type core,
\begin{equation}\label{small-r-f}
 f(r)=1-\frac{2M}{l_0^3}r^2
 +\left(\frac{3M}{l_0^5}-\frac{M^2}{5l_0^6}\right)r^4
 +\Order{r^6},
\end{equation}
so the curvature invariants remain finite at $r=0$.  Horizons are located at the positive roots of $f(r)=0$.  For $M=1$, the two horizons merge at the extremal value $l_0\simeq0.8581$; all values of $l_0$ used below are below this value and therefore describe black holes with a regular event horizon.

\begin{figure}[htb!]
\centering
\includegraphics[width=\columnwidth]{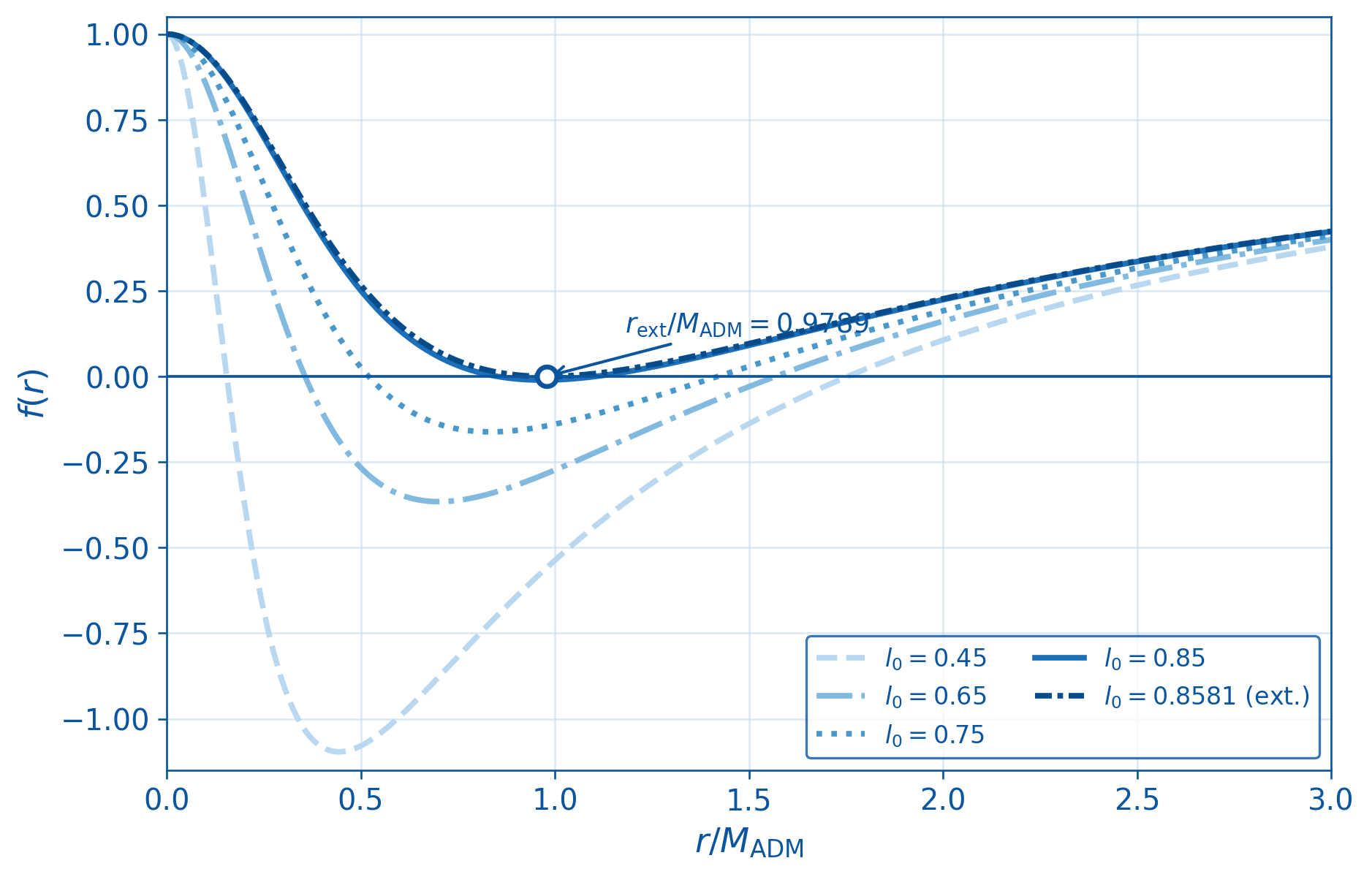}
\caption{Metric function $f(r)$ for $M=1$ and representative zero-point lengths, including the extremal value $l_0=0.8580937$.  The radial coordinate is normalized by $M_{\rm ADM}=M+3\pi M^2/(32l_0)$.  Each subextremal curve has a Cauchy and an event horizon, whereas the extremal curve is tangent to $f=0$ at $r_{\rm ext}/M_{\rm ADM}=0.9788831$, as indicated by the circle.}\label{fig:metric-function}
\end{figure}

Figure~\ref{fig:metric-function} displays the resulting metric function in ADM-scaled coordinates for representative values used in our analysis, together with the extremal curve at $l_0=0.8580937$.  The four subextremal curves cross zero twice, corresponding to the inner and outer horizons.  The extremal curve instead touches zero at $r_{\rm ext}=1.3148661$, or $r_{\rm ext}/M_{\rm ADM}=0.9788831$, where the two roots form a double horizon.

The axial gravitational perturbations considered in this work reduce to a Regge-Wheeler-type equation. We use the Regge--Wheeler gauge~\cite{Regge:1957td} and separate the time and angular dependence as $e^{-\imo\omega t}$ and axial vector spherical harmonics.  Because the background is spherically symmetric, it is sufficient to display the axisymmetric representative, for which
\begin{equation}\label{axial-metric-ansatz}
 \begin{split}
 \delta g_{t\phi}=\delta g_{\phi t}
 &=h_0(r)e^{-\imo\omega t}X_\ell(\theta),\\
 \delta g_{r\phi}=\delta g_{\phi r}
 &=h_1(r)e^{-\imo\omega t}X_\ell(\theta),\\
 X_\ell(\theta)&=\sin\theta\,
 \frac{dP_\ell(\cos\theta)}{d\theta},
 \end{split}
\end{equation}
For arbitrary azimuthal number, $X_\ell$ is replaced by the corresponding axial vector harmonic, without changing the radial equations.

The quantum-corrected background is represented by the effective anisotropic tensor
$T_{\mu\nu}=(\rho+p_t)u_\mu u_\nu+p_tg_{\mu\nu}+(p_r-p_t)s_\mu s_\nu$, with $u^\mu=(f^{-1/2},0,0,0)$ and $s^\mu=(0,f^{1/2},0,0)$~\cite{Ashtekar:2018lag,Ashtekar:2018cay}.  Since $\rho$, $p_r$, and $p_t$ are rotational scalars, their odd-parity perturbations vanish.  The possible axial vector perturbations can be separated as~\cite{Bouhmadi-Lopez:2020oia,Konoplya:2024lch}
\begin{eqnarray}\label{axial-fluid-perturbations}
 \delta u_\phi&=&-\imo\omega U(r)e^{-\imo\omega t}X_\ell(\theta),\nonumber\\
 \delta s_\phi&=&-S(r)e^{-\imo\omega t}X_\ell(\theta).
\end{eqnarray}
The approximation used here is precisely $\delta s^\mu=0$, or $S(r)=0$.  With this closure, the linearized conservation equation $\delta(\nabla_\mu T^{\mu\nu})=0$ sets the remaining axial velocity amplitude to $U(r)=0$.  Thus the odd-parity matter variables do not supply an independent dynamical degree of freedom, and the $t\phi$ and $r\phi$ components of the linearized Einstein equations form the closed radial system
\begin{eqnarray}\label{axial-first-order-system}
 0&=&\left[r^2\omega^2-(\ell-1)(\ell+2)f(r)\right]h_1(r)
 \nonumber\\
 &&-\imo r^2\omega h_0'(r)+2\imo r\omega h_0(r),\nonumber\\
 0&=&f'(r)h_1(r)+2f(r)h_1'(r)
 +\frac{2\imo\omega}{f(r)}h_0(r) .
\end{eqnarray}
Eliminating $h_0$ and defining
\begin{equation}\label{tortoise}
 h_1(r)=\frac{r}{f(r)}\Psi(r),
 \qquad
 \frac{dr_*}{dr}=\frac{1}{f(r)},
\end{equation}
reduces Eq.~(\ref{axial-first-order-system}) to
\begin{equation}\label{wave-equation}
 \frac{d^2\Psi}{dr_*^2}+\left[\omega^2-V_\ell(r)\right]\Psi=0 .
\end{equation}
For the effective anisotropic source, the axial potential can be written as (see the derivations \cite{Bouhmadi-Lopez:2020oia,Konoplya:2025hgp,Konoplya:2024lch,Bolokhov:2025egl})
\begin{eqnarray}\label{grav-potential}
 V_\ell(r)&=&f(r)\left[\frac{\ell(\ell+1)}{r^2} -\frac{6m(r)}{r^3}+4\pi\left(\rho-p_r\right)\right]\nonumber\\
 &=&\frac{f(r)}{r^2}\left[\ell(\ell+1)-2+2f(r)-rf'(r)\right].
\end{eqnarray}
This derivation is the direct specialization of Refs.~\cite{Bouhmadi-Lopez:2020oia,Konoplya:2024lch} to $g_{tt}=-1/g_{rr}=-f(r)$; analogous axial decouplings appear in Refs.~\cite{Bronnikov:2012ch,Chen:2019iuo,Chakraborty:2024gcr,Bolokhov:2025lnt,Lutfuoglu:2026rqe,Lutfuoglu:2026zel}.  Its limitation is also explicit: setting $S=U=0$ selects the metric-led axial branch.  If the effective anisotropic medium has independent axial dynamics, $S$ and $U$ must be retained and supplemented by constitutive equations, leading to a coupled matter--gravity eigenvalue problem rather than the single potential in Eq.~(\ref{grav-potential})~\cite{Zhao:2026eti}. In the limit $l_0\to0$ with the ADM mass kept fixed, this potential reduces to the standard Schwarzschild Regge-Wheeler potential.  Throughout the numerical calculations of quasinormal modes, all dimensional quantities were measured in units of $M$, that is, supposing $M=1$. However, the rescaling to units of ADM mass is done afterwards.

\begin{figure*}
\centering
\includegraphics[width=0.95\textwidth]{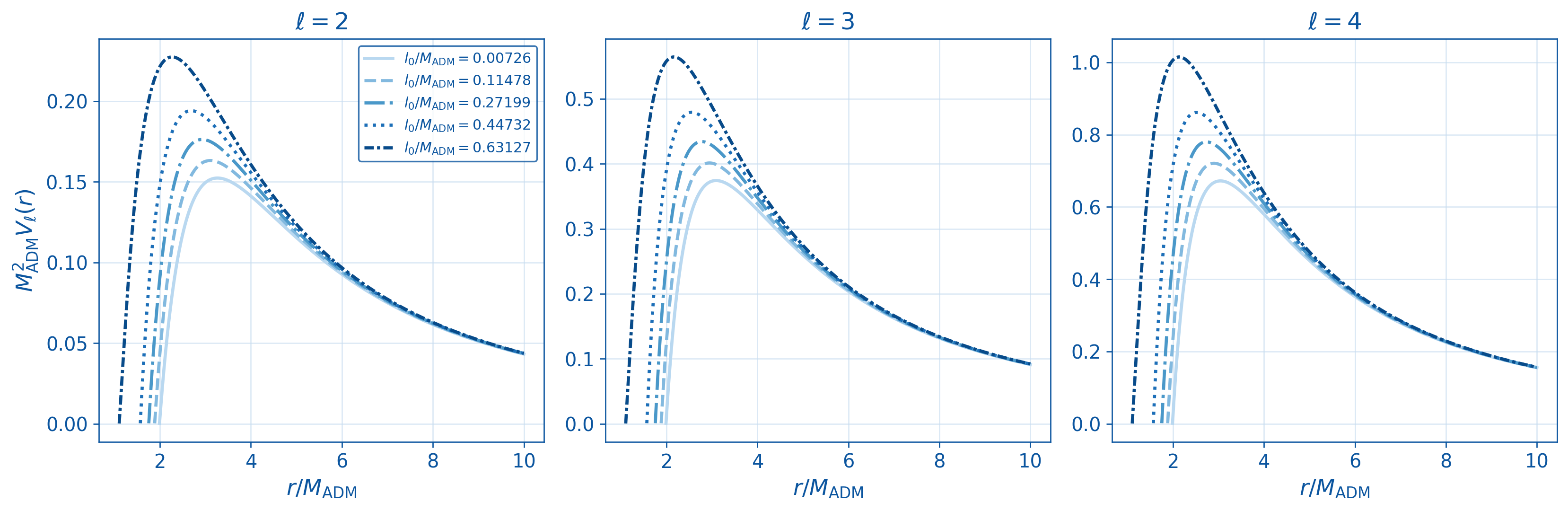}
\caption{Effective axial gravitational potentials for the regular self-energy black hole.  The three panels show $\ell=2,3,4$, while the curves correspond to $l_0/M_{\rm ADM}=0.00726$, $0.11478$, $0.27199$, $0.44732$, and $0.63127$ (equivalently $l_0=0.05$, $0.25$, $0.45$, $0.65$, and $0.85$ for $M=1$).}
\label{fig:grav-effective-potentials}
\end{figure*}

The dependence of the axial barrier on the zero-point length and on the multipole number is illustrated in Fig.~\ref{fig:grav-effective-potentials}.  The potentials are plotted in ADM-scaled variables, $M_{\rm ADM}^2V_\ell$ versus $r/M_{\rm ADM}$, for representative values of $l_0$ used in the frequency tables.  For fixed $\ell$, increasing $l_0/M_{\rm ADM}$ raises the height of the barrier and shifts its maximum toward smaller $r/M_{\rm ADM}$.  At fixed $l_0/M_{\rm ADM}$, larger multipoles produce the expected higher and narrower Regge--Wheeler barrier.  The largest displayed value is the near-extremal case $l_0=0.85$, corresponding to $l_0/M_{\rm ADM}=0.63127$ for $M=1$.

\section{Methods for quasinormal modes}\label{sec:methods}

The numerical analysis combines a frequency-domain calculation with a direct evolution of the master equation.  The tabulated spectra are obtained from a high-order WKB expansion improved by Pad\'e resummation.  This approach is particularly efficient for the axial gravitational potential considered here, because the potential has the usual single-barrier shape illustrated in Fig.~\ref{fig:grav-effective-potentials}.  The time-domain calculation is used as a separate diagnostic: it evolves the same wave equation without expanding around the potential maximum and then extracts the dominant ringing frequency from the waveform.

\subsection{Higher-order WKB method}\label{subsec:wkb}

We start from the one-dimensional problem in Eq.~(\ref{wave-equation}).  With the time dependence chosen as $e^{-\imo\omega t}$, quasinormal modes are defined by excluding incoming radiation from both asymptotic boundaries.  Thus, the radial function behaves as
\begin{equation}\label{boundaryconditions}
\Psi(r_*\to\pm\infty)\propto e^{\pm\imo \omega r_*},
\end{equation}
where the lower sign gives a wave falling through the event horizon and the upper sign gives a wave escaping to spatial infinity.  The WKB method implements these boundary conditions by expanding the potential near its maximum and connecting the local solution across the two asymptotic regions.  If $r_0$ is the position of the maximum, we denote by $V_0$ the value of the potential at $r_0$ and by $V_2$ its second derivative with respect to $r_*$ at the same point.  In the notation of Ref.~\cite{Konoplya:2019hlu}, the quantization condition may be written in the form
\begin{eqnarray}\label{WKBformula-spherical}
\omega^2&=&V_0+A_2(\K^2)+A_4(\K^2)+A_6(\K^2)+\ldots\nonumber\\
&&-\imo \K\sqrt{-2V_2}\left(1+A_3(\K^2)+A_5(\K^2)\right.\nonumber\\
&&\left.+A_7(\K^2)+\ldots\right),
\end{eqnarray}
with
\begin{equation}
\K=n+\frac{1}{2}, \qquad n=0,1,2,\ldots .
\end{equation}
Here $n$ is the overtone number.  The correction $A_i$ is a polynomial expression built from $\K$ and from derivatives of $V_\ell(r)$ at the peak up to order $2i$.  The second- and third-order terms were obtained in Ref.~\cite{Iyer:1986np}, the fourth- to sixth-order contributions in Refs.~\cite{Konoplya:2003ii}, and the extension through very high orders in Refs.~\cite{Matyjasek:2017psv,Matyjasek:2019eeu,Hatsuda:2019eoj,Matyjasek:2026yiu,Konoplya:2026rjh}.  The same WKB machinery, with different truncations and improvements, has been used in a broad range of black-hole spectral and scattering problems; representative applications include Refs.~\cite{Dubinsky:2024hmn,Konoplya:2009hv,Konoplya:2023moy,Lutfuoglu:2025kqp,Konoplya:2010vz,Malik:2026jzl,Dubinsky:2024rvf,Kodama:2009bf,Malik:2024tuf,Fernando:2016ftj,Guo:2020blq,Albuquerque:2023lhm,Malik:2023bxc,Konoplya:2007yy,Dubinsky:2025ypj,Dubinsky:2025wns,Konoplya:2002wt,Tan:2022vfe,Kokkotas:2010zd,Lutfuoglu:2026zel,Konoplya:2019ppy}.

A direct truncation of the WKB series is not always the most stable way to use the high-order information.  We therefore introduce a formal expansion parameter $\varepsilon$, construct the WKB series in powers of $\varepsilon$, and replace it by the rational Pad\'e approximant \cite{Matyjasek:2017psv}
\begin{equation}
P_{\tilde m/\tilde n}(\varepsilon)=
\frac{\sum_{i=0}^{\tilde m} a_i \varepsilon^i}
     {\sum_{j=0}^{\tilde n} b_j \varepsilon^j}.
\end{equation}
The physical value is obtained after setting $\varepsilon=1$.  The integers $\tilde m$ and $\tilde n$ specify the numerator and denominator degrees, with $\tilde m+\tilde n$ equal to the WKB order being resummed~\cite{Konoplya:2019hlu}.  In practice, we use the 16th- and 14th-order WKB results and compare the symmetric Pad\'e choices $\tilde m=8$ and $\tilde m=7$, respectively.  This comparison gives an internal estimate of numerical stability.  Symmetric Pad\'e approximants have proved especially accurate in recent applications~\cite{Lutfuoglu:2026boa,Skvortsova:2026jtx,Bolokhov:2026uol,Bolokhov:2026dfg,Lutfuoglu:2026rqe,Skvortsova:2026idf,Lutfuoglu:2026zxj}, and we adopt the same prescription here.

\subsection{Time-domain integration and Prony analysis}\label{subsec:timedomain}

As an independent check, we also solve Eq.~(\ref{wave-equation}) as an initial-value problem.  It is convenient to introduce the null coordinates $u=t-r_*$ and $v=t+r_*$.  On the corresponding characteristic grid, the value of the field at the future vertex of a null cell is advanced from the other three vertices by the Gundlach-Price-Pullin integration scheme~\cite{Gundlach:1993tp},
\begin{eqnarray}
\Psi\left(N\right)&=&\Psi\left(W\right)+\Psi\left(E\right)-\Psi\left(S\right)\nonumber\\
&&- \Delta^2V\left(S\right)\frac{\Psi\left(W\right)+\Psi\left(E\right)}{8}+{\cal O}\left(\Delta^4\right),\label{Discretization}
\end{eqnarray}
where $S=(u,v)$, $W=(u+\Delta,v)$, $E=(u,v+\Delta)$ and $N=(u+\Delta,v+\Delta)$.  This characteristic scheme has become a standard tool for testing black-hole ringing and late-time behavior in many backgrounds~\cite{Momennia:2022tug,Skvortsova:2023zca,Aneesh:2018hlp,Abdalla:2012si,Konoplya:2014lha,Dubinsky:2024jqi,Konoplya:2006gq,Bolokhov:2024ixe,Malik:2024iky,Konoplya:2013sba,Bolokhov:2024bke,Lutfuoglu:2025pzi,Dubinsky:2025nxv,Konoplya:2018yrp,Malik:2024bmp,Dubinsky:2024gwo,Lutfuoglu:2026fpx,Konoplya:2005et,Konoplya:2023fmh,Dubinsky:2025bvf}.

The initial data are chosen so that the early part of the signal contains a prompt response followed by a clean exponentially damped ringing stage.  The frequency is then extracted only from a time interval where this ringing dominates and before the late-time tail becomes important.  In that interval, the waveform is approximated by a finite Prony sum,
\begin{equation}
\Psi(t)\simeq \sum_{j=1}^{p} C_j e^{-\imo\omega_j t}.
\end{equation}
The fitted complex frequencies are monitored under changes of the fitting window and of the number of exponents $p$.  We use this procedure only as a consistency check of the fundamental modes; the WKB--Pad\'e calculation remains the source of the tabulated values.

\section{Quasinormal modes and time-domain evolution}\label{sec:qnm}

\begin{figure*}
\centering
\includegraphics[width=0.95\textwidth]{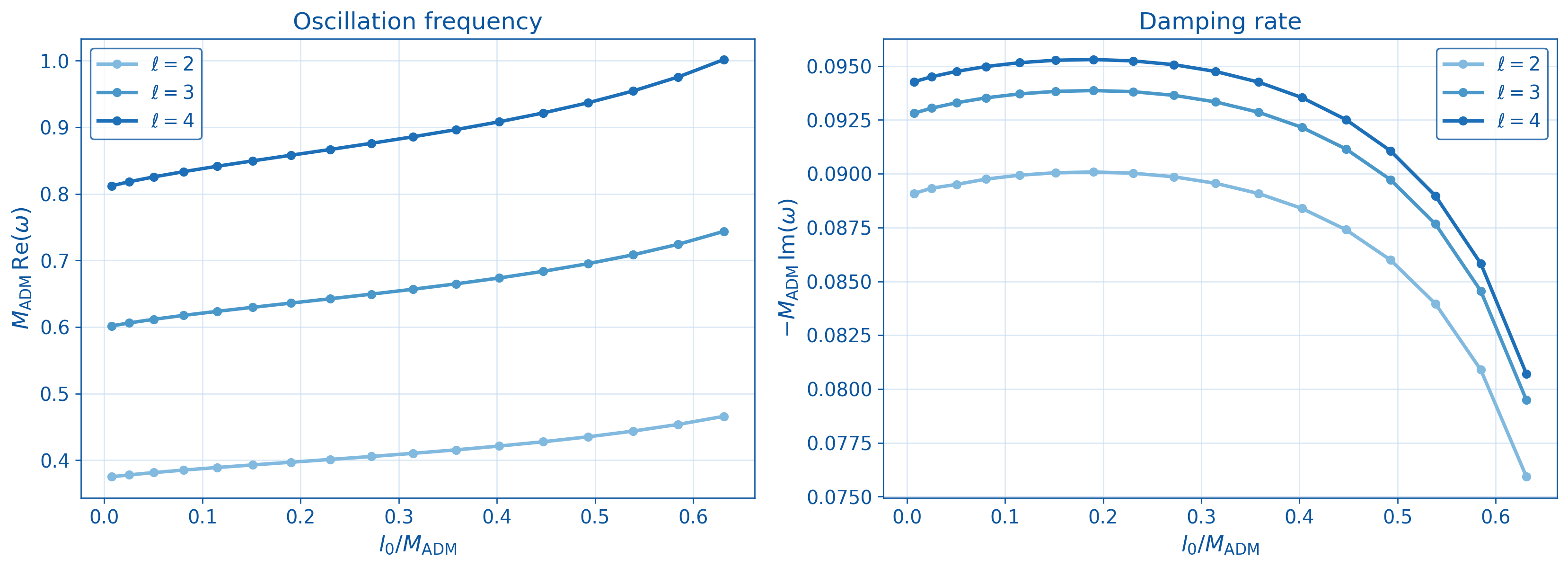}
\caption{Fundamental axial gravitational quasinormal modes in ADM units for zero-point lengths up to $l_0=0.85$ ($l_0/M_{\rm ADM}=0.63127$).  The left panel shows $M_{\rm ADM}\mathrm{Re}(\omega)$, while the right panel shows the damping rate $-M_{\rm ADM}\mathrm{Im}(\omega)$, both as functions of $l_0/M_{\rm ADM}$.}
\label{fig:qnm-adm-fundamental}
\end{figure*}

\begin{figure*}
\centering
\includegraphics[width=0.95\textwidth]{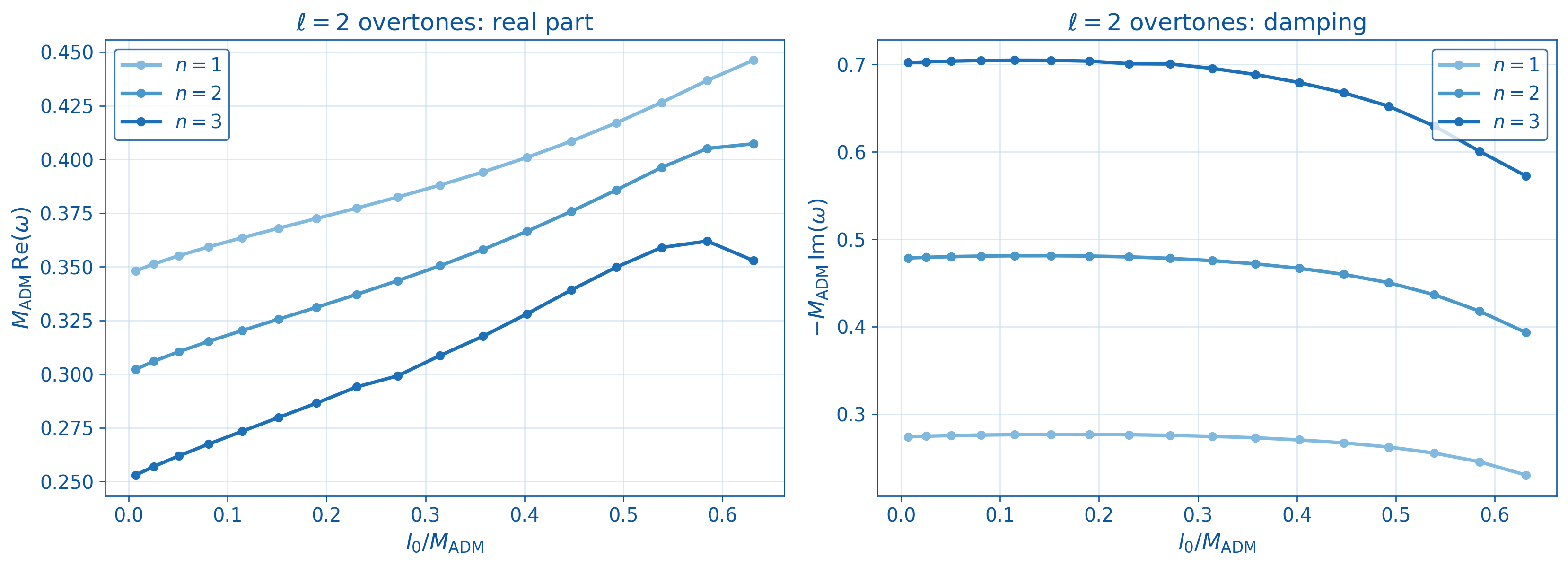}
\caption{First three overtones of the $\ell=2$ axial gravitational mode in ADM units for zero-point lengths up to $l_0=0.85$.  The left panel gives $M_{\rm ADM}\mathrm{Re}(\omega)$ and the right panel gives $-M_{\rm ADM}\mathrm{Im}(\omega)$ as functions of $l_0/M_{\rm ADM}$.}
\label{fig:qnm-adm-overtones}
\end{figure*}

The tables in this section list the WKB--Pad\'e frequencies in units of the parameter $M$, which was set equal to unity in the numerical integration.  Since the physical asymptotic mass of the self-energy geometry is instead $M_{\rm ADM}=1+3\pi/(32l_0)$, it is also useful to present the same spectra in ADM units.  In Figs.~\ref{fig:qnm-adm-fundamental} and \ref{fig:qnm-adm-overtones} we therefore plot $M_{\rm ADM}\omega$ as a function of the dimensionless ratio $l_0/M_{\rm ADM}$.  This rescaling removes the large change of the total mass at small $l_0$ and makes the deformation of the spectrum by the zero-point length more transparent.

For the fundamental modes shown in Fig.~\ref{fig:qnm-adm-fundamental}, the oscillation frequency grows monotonically with $l_0/M_{\rm ADM}$ for all three multipoles up to the near-extremal point $l_0=0.85$.  The damping rate $-\mathrm{Im}(M_{\rm ADM}\omega)$ behaves differently: it has a shallow maximum near $l_0=0.35$ and then decreases increasingly rapidly toward the near-extremal endpoint.  Thus, in ADM units, the self-energy correction makes the fundamental gravitational modes oscillate faster, while the damping is reduced in the region closest to extremality.  The hierarchy with respect to $\ell$ remains the expected one: larger $\ell$ gives a larger real part and a slightly larger damping rate.

This trend is consistent with the potential profiles in Fig.~\ref{fig:grav-effective-potentials}: in the WKB picture, the real part of the QNM frequency is controlled mainly by the barrier height $V_0$, so the higher barriers produced by larger $l_0/M_{\rm ADM}$ naturally lead to larger oscillation frequencies.

The damping rate is controlled by a different combination of barrier properties.  At leading WKB order,
\begin{equation}
 -{\rm Im}(\omega)\simeq
 \left(n+\frac{1}{2}\right)
 \frac{\sqrt{-2V_0^{(2)}}}{2\sqrt{V_0}},
\end{equation}
where $V_0^{(2)}$ is the second derivative with respect to $r_*$ at the maximum.  Although $V_0$ grows monotonically with $l_0/M_{\rm ADM}$, the curvature-to-height ratio in this expression first increases slightly and then decreases.  As the two horizons approach degeneracy, the surface gravity decreases and the barrier becomes broader in the tortoise coordinate, reducing its normalized curvature.  The competition between the increasing height and the subsequent broadening therefore explains why the fundamental oscillation frequency remains monotonic while its damping rate develops a maximum and then decreases.

The overtone data for $\ell=2$ are summarized in Fig.~\ref{fig:qnm-adm-overtones}.  After ADM rescaling, the real parts of the $n=1$ and $n=2$ modes increase monotonically up to $l_0=0.85$, whereas the $n=3$ real part reaches its largest tabulated value at $l_0=0.80$ and turns over at $l_0=0.85$.  The damping rates remain ordered by overtone number; they reach shallow maxima at small or intermediate $l_0/M_{\rm ADM}$ and then decrease toward the near-extremal endpoint.  The close agreement between the 16th- and 14th-order Pad\'e values in the tables indicates that these trends are not artifacts of the WKB order.

The turnover of the $n=3$ real part is likewise associated with the near-extremal change in the barrier shape rather than with its height alone.  Higher overtones depend more strongly on the higher derivatives entering the WKB corrections and probe a wider portion of the potential, including the increasingly stretched near-horizon region.  Near $l_0=0.85$, these shape corrections become large enough to overcome the upward shift produced by the increasing peak height.  The effect is unlikely to be a WKB truncation artifact: the ADM-scaled $n=3$ real part decreases by about $2.5\%$ between $l_0=0.80$ and $0.85$, whereas the difference between the two Pad\'e orders at $l_0=0.85$ is about $0.16\%$.  The potential remains a smooth single barrier and all imaginary parts remain negative, so this behavior does not indicate an instability; nevertheless, an independent continued-fraction calculation would be a useful quantitative check.

For completeness, the numerical values used to construct these plots are displayed in Tables~I and II in the original $M=1$ normalization.  The last column gives the relative difference between the two Pad\'e-resummed WKB orders and provides a compact estimate of the numerical stability.

\begin{table}
\centering
\scriptsize
\setlength{\tabcolsep}{3.5pt}
\renewcommand{\arraystretch}{0.88}
\begin{tabular}{c c c c}
\hline
$l_0$ & WKB16 ($\tilde{m}=8$) & WKB14 ($\tilde{m}=7$) & difference  \\
\hline
\multicolumn{4}{c}{$\ell=2$}\\
\hline
$0.05$ & $0.054430-0.012929 i$ & $0.054430-0.012928 i$ & $0.0029\%$\\
$0.1$ & $0.095800-0.022642 i$ & $0.095798-0.022635 i$ & $0.0071\%$\\
$0.15$ & $0.128715-0.030203 i$ & $0.128713-0.030216 i$ & $0.0098\%$\\
$0.2$ & $0.155776-0.036299 i$ & $0.155774-0.036303 i$ & $0.0024\%$\\
$0.25$ & $0.178600-0.041290 i$ & $0.178599-0.041291 i$ & $0.0010\%$\\
$0.3$ & $0.198272-0.045438 i$ & $0.198272-0.045439 i$ & $0.0004\%$\\
$0.35$ & $0.215560-0.048918 i$ & $0.215560-0.048918 i$ & $0.0001\%$\\
$0.4$ & $0.231034-0.051849 i$ & $0.231034-0.051849 i$ & $0\%$\\
$0.45$ & $0.245137-0.054312 i$ & $0.245137-0.054312 i$ & $0\%$\\
$0.5$ & $0.258228-0.056358 i$ & $0.258228-0.056358 i$ & $0\%$\\
$0.55$ & $0.270612-0.058016 i$ & $0.270612-0.058016 i$ & $0\%$\\
$0.6$ & $0.282564-0.059288 i$ & $0.282564-0.059288 i$ & $0\%$\\
$0.65$ & $0.294347-0.060148 i$ & $0.294347-0.060148 i$ & $0.\times 10^{\text{-4}}\%$\\
$0.7$ & $0.306238-0.060529 i$ & $0.306238-0.060529 i$ & $0\%$\\
$0.75$ & $0.318550-0.060286 i$ & $0.318550-0.060286 i$ & $0.00002\%$\\
$0.8$ & $0.331663-0.059125 i$ & $0.331663-0.059125 i$ & $0\%$\\
$0.85$ & $0.346011-0.056383 i$ & $0.346011-0.056383 i$ & $0.\times 10^{\text{-4}}\%$\\
\hline
\multicolumn{4}{c}{$\ell=3$}\\
\hline
$0.05$ & $0.087310-0.013470 i$ & $0.087310-0.013470 i$ & $0\%$\\
$0.1$ & $0.153646-0.023585 i$ & $0.153646-0.023585 i$ & $0\%$\\
$0.15$ & $0.206400-0.031483 i$ & $0.206400-0.031483 i$ & $0\%$\\
$0.2$ & $0.249742-0.037826 i$ & $0.249742-0.037826 i$ & $0\%$\\
$0.25$ & $0.286273-0.043024 i$ & $0.286273-0.043024 i$ & $0\%$\\
$0.3$ & $0.317730-0.047346 i$ & $0.317730-0.047346 i$ & $0\%$\\
$0.35$ & $0.345345-0.050973 i$ & $0.345345-0.050973 i$ & $0\%$\\
$0.4$ & $0.370030-0.054030 i$ & $0.370030-0.054030 i$ & $0\%$\\
$0.45$ & $0.392494-0.056600 i$ & $0.392494-0.056600 i$ & $0\%$\\
$0.5$ & $0.413309-0.058738 i$ & $0.413309-0.058738 i$ & $0\%$\\
$0.55$ & $0.432960-0.060475 i$ & $0.432960-0.060475 i$ & $0\%$\\
$0.6$ & $0.451884-0.061814 i$ & $0.451884-0.061814 i$ & $0\%$\\
$0.65$ & $0.470502-0.062730 i$ & $0.470502-0.062730 i$ & $0\%$\\
$0.7$ & $0.489256-0.063155 i$ & $0.489256-0.063155 i$ & $0\%$\\
$0.75$ & $0.508657-0.062946 i$ & $0.508657-0.062946 i$ & $0\%$\\
$0.8$ & $0.529374-0.061803 i$ & $0.529374-0.061803 i$ & $0\%$\\
$0.85$ & $0.552357-0.059028 i$ & $0.552357-0.059028 i$ & $0\%$\\
\hline
\multicolumn{4}{c}{$\ell=4$}\\
\hline
$0.05$ & $0.117855-0.013682 i$ & $0.117855-0.013682 i$ & $0\%$\\
$0.1$ & $0.207386-0.023955 i$ & $0.207386-0.023955 i$ & $0\%$\\
$0.15$ & $0.278569-0.031975 i$ & $0.278569-0.031975 i$ & $0\%$\\
$0.2$ & $0.337039-0.038414 i$ & $0.337039-0.038414 i$ & $0\%$\\
$0.25$ & $0.386304-0.043690 i$ & $0.386304-0.043690 i$ & $0\%$\\
$0.3$ & $0.428712-0.048076 i$ & $0.428712-0.048076 i$ & $0\%$\\
$0.35$ & $0.465924-0.051755 i$ & $0.465924-0.051755 i$ & $0\%$\\
$0.4$ & $0.499171-0.054855 i$ & $0.499171-0.054855 i$ & $0\%$\\
$0.45$ & $0.529408-0.057461 i$ & $0.529408-0.057461 i$ & $0\%$\\
$0.5$ & $0.557405-0.059628 i$ & $0.557405-0.059628 i$ & $0\%$\\
$0.55$ & $0.583815-0.061387 i$ & $0.583815-0.061387 i$ & $0\%$\\
$0.6$ & $0.609223-0.062742 i$ & $0.609223-0.062742 i$ & $0\%$\\
$0.65$ & $0.634196-0.063668 i$ & $0.634196-0.063668 i$ & $0\%$\\
$0.7$ & $0.659327-0.064098 i$ & $0.659327-0.064098 i$ & $0\%$\\
$0.75$ & $0.685310-0.063886 i$ & $0.685310-0.063886 i$ & $0\%$\\
$0.8$ & $0.713058-0.062734 i$ & $0.713058-0.062734 i$ & $0\%$\\
$0.85$ & $0.743949-0.059933 i$ & $0.743949-0.059933 i$ & $0\%$\\
\hline
\end{tabular}
\caption{Fundamental QNMs of the $\ell=2,3,4$ perturbations of the Jusufi--Singleton black hole ($M=1$), calculated using the WKB formula at different orders and Pad\'e approximants.}
\end{table}

\begin{figure*}[t]
\centering
\textbf{(a)}\\[-1mm]
\includegraphics[width=0.72\textwidth]{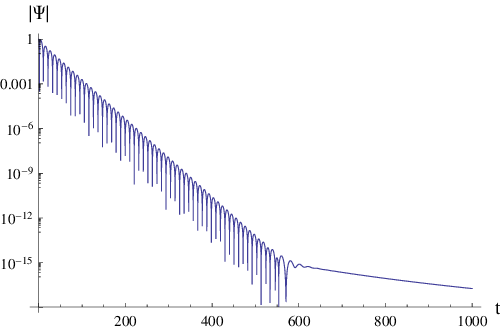}\\[1mm]
\textbf{(b)}\\[-1mm]
\includegraphics[width=0.72\textwidth]{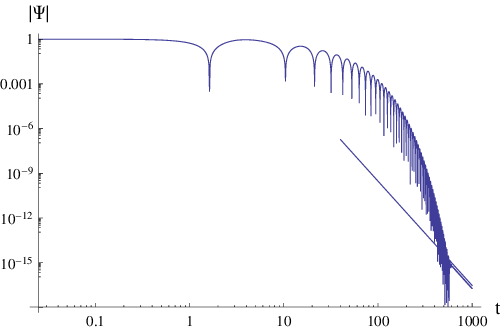}
\caption{Time-domain profile of the axial gravitational perturbation for $\ell=2$ and $l_0=0.65$.  Panel (a) shows the semi-logarithmic ringdown plot used for the Prony extraction, which gives $\omega_{\rm TD}=0.29435-0.06015\imo$.  Panel (b) shows the same evolution on a log-log scale; the straight late-time segment follows $t^{-7}$, as expected from Price's law for the quadrupole gravitational tail.}
\label{fig:td-grav-l2-l065}
\end{figure*}

\begin{figure*}[t]
\centering
\textbf{(a)}\\[-1mm]
\includegraphics[width=0.72\textwidth]{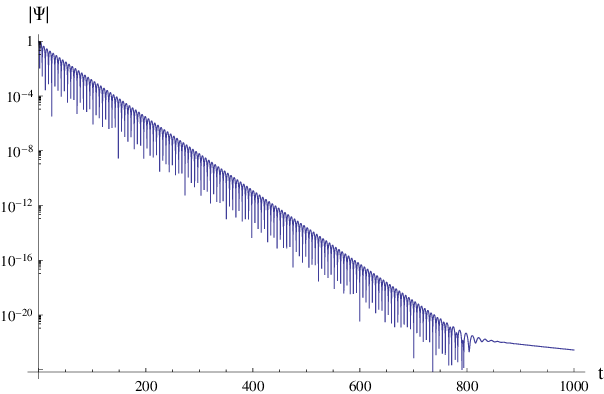}\\[1mm]
\textbf{(b)}\\[-1mm]
\includegraphics[width=0.72\textwidth]{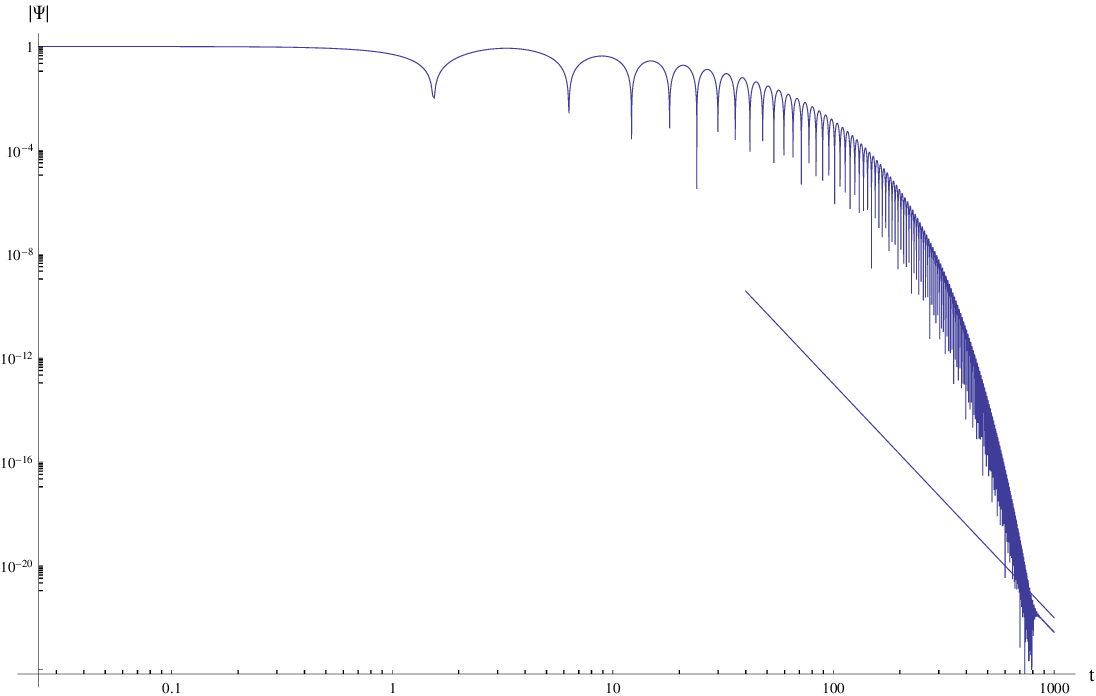}
\caption{Time-domain profile of the axial gravitational perturbation for $\ell=3$ and $l_0=0.8$.  Panel (a) shows the semi-logarithmic ringdown plot used for the Prony extraction, which gives $\omega_{\rm TD}=0.5294-0.0618\imo$.  Panel (b) shows the same evolution on a log-log scale; the straight late-time segment follows $t^{-9}$, as expected from Price's law for the $\ell=3$ gravitational tail.}
\label{fig:td-grav-l3-l080}
\end{figure*}

The time-domain signals in Figs.~\ref{fig:td-grav-l2-l065} and \ref{fig:td-grav-l3-l080} provide independent checks of both the ringdown frequencies and the asymptotic decay.  For $\ell=2$ and $l_0=0.65$, the WKB--Pad\'e value is $\omega_{\rm WKB}=0.294347-0.060148\imo$, while the Prony extraction gives $\omega_{\rm TD}=0.29435-0.06015\imo$.  For $\ell=3$ and $l_0=0.8$, the corresponding values are $\omega_{\rm WKB}=0.529374-0.061803\imo$ and $\omega_{\rm TD}=0.5294-0.0618\imo$.  Using the complex relative difference
\begin{equation}
 \Delta_{\rm TD/WKB}=\frac{|\omega_{\rm TD}-\omega_{\rm WKB}|}{|\omega_{\rm WKB}|}\times100\%,
\end{equation}
one obtains $\Delta_{\rm TD/WKB}=1.20\times10^{-3}\%$ for the $\ell=2$ profile and $4.91\times10^{-3}\%$ for the $\ell=3$ profile.  The corresponding relative differences between the damping rates are $3.33\times10^{-3}\%$ and $4.85\times10^{-3}\%$, respectively.  This agreement is consistent with the expected accuracy of the comparison: the WKB--Pad\'e values are very stable for the fundamental modes, whereas the time-domain estimates are limited mainly by the choice of the Prony fitting window, residual contamination from the prompt response, and the onset of the late-time tail.

At later times, the logarithmic plot in Fig.~\ref{fig:td-grav-l2-l065} (b) displays a straight asymptotic segment proportional to $t^{-7}$.  This is exactly the Price-law exponent $t^{-(2\ell+3)}$ for a massless perturbation with compact initial data at fixed radius when $\ell=2$~\cite{Price:1972pw}. A similar result for $\ell=3$ and $l_0=0.8$ is shown in Fig. \ref{fig:td-grav-l3-l080}. The result is physically expected in the present geometry because the spacetime is asymptotically flat and the axial gravitational potential has the same leading centrifugal term and the same long-range $1/r^3$ structure that controls the Schwarzschild tail, while the zero-point-length corrections enter only through subleading inverse powers of $r$.  Therefore, the numerical evolution confirms that the self-energy regularization changes the quasinormal ringing frequencies but does not alter the leading quadrupole Price-law tail in the asymptotic regime shown here.

\begin{table}
\centering
\scriptsize
\setlength{\tabcolsep}{3.5pt}
\renewcommand{\arraystretch}{0.88}
\begin{tabular}{c c c c}
\hline
$l_0$ & WKB16 ($\tilde{m}=8$) & WKB14 ($\tilde{m}=7$) & difference  \\
\hline
\multicolumn{4}{c}{$n=1$}\\
\hline
$0.05$ & $0.050516-0.039794 i$ & $0.050542-0.039830 i$ & $0.0683\%$\\
$0.1$ & $0.089037-0.069656 i$ & $0.089076-0.069707 i$ & $0.0567\%$\\
$0.15$ & $0.119833-0.092955 i$ & $0.119874-0.093003 i$ & $0.0413\%$\\
$0.2$ & $0.145291-0.111646 i$ & $0.145324-0.111681 i$ & $0.0263\%$\\
$0.25$ & $0.166897-0.126944 i$ & $0.166920-0.126965 i$ & $0.0146\%$\\
$0.3$ & $0.185654-0.139640 i$ & $0.185666-0.139652 i$ & $0.0072\%$\\
$0.35$ & $0.202275-0.150267 i$ & $0.202283-0.150274 i$ & $0.0044\%$\\
$0.4$ & $0.217297-0.159194 i$ & $0.217306-0.159203 i$ & $0.0046\%$\\
$0.45$ & $0.231153-0.166658 i$ & $0.231153-0.166662 i$ & $0.0017\%$\\
$0.5$ & $0.244182-0.172812 i$ & $0.244189-0.172813 i$ & $0.0024\%$\\
$0.55$ & $0.256674-0.177737 i$ & $0.256681-0.177734 i$ & $0.0023\%$\\
$0.6$ & $0.268900-0.181436 i$ & $0.268904-0.181433 i$ & $0.0014\%$\\
$0.65$ & $0.281101-0.183813 i$ & $0.281100-0.183814 i$ & $0.0003\%$\\
$0.7$ & $0.293491-0.184645 i$ & $0.293491-0.184646 i$ & $0.0003\%$\\
$0.75$ & $0.306238-0.183470 i$ & $0.306237-0.183469 i$ & $0.0004\%$\\
$0.8$ & $0.319250-0.179404 i$ & $0.319250-0.179403 i$ & $0.0002\%$\\
$0.85$ & $0.331359-0.171073 i$ & $0.331360-0.171072 i$ & $0.0005\%$\\
\hline
\multicolumn{4}{c}{$n=2$}\\
\hline
$0.05$ & $0.043869-0.069462 i$ & $0.044109-0.069188 i$ & $0.444\%$\\
$0.1$ & $0.077552-0.121504 i$ & $0.077838-0.121171 i$ & $0.304\%$\\
$0.15$ & $0.104751-0.162017 i$ & $0.104993-0.161735 i$ & $0.193\%$\\
$0.2$ & $0.127497-0.194441 i$ & $0.127678-0.194234 i$ & $0.118\%$\\
$0.25$ & $0.147055-0.220909 i$ & $0.147186-0.220767 i$ & $0.0728\%$\\
$0.3$ & $0.164284-0.242814 i$ & $0.164391-0.242706 i$ & $0.0519\%$\\
$0.35$ & $0.179812-0.261101 i$ & $0.179966-0.260976 i$ & $0.0625\%$\\
$0.4$ & $0.194156-0.276379 i$ & $0.194226-0.276442 i$ & $0.0278\%$\\
$0.45$ & $0.207631-0.289012 i$ & $0.207647-0.289059 i$ & $0.0139\%$\\
$0.5$ & $0.220546-0.299283 i$ & $0.220584-0.299410 i$ & $0.0356\%$\\
$0.55$ & $0.233189-0.307325 i$ & $0.232827-0.307398 i$ & $0.0957\%$\\
$0.6$ & $0.245821-0.313109 i$ & $0.245784-0.313061 i$ & $0.0152\%$\\
$0.65$ & $0.258613-0.316456 i$ & $0.258613-0.316455 i$ & $0.0004\%$\\
$0.7$ & $0.271489-0.316960 i$ & $0.271619-0.316936 i$ & $0.0316\%$\\
$0.75$ & $0.284490-0.313592 i$ & $0.284555-0.313649 i$ & $0.0205\%$\\
$0.8$ & $0.296084-0.305282 i$ & $0.296107-0.305296 i$ & $0.0063\%$\\
$0.85$ & $0.302457-0.292149 i$ & $0.302437-0.292232 i$ & $0.0201\%$\\
\hline
\multicolumn{4}{c}{$n=3$}\\
\hline
$0.05$ & $0.036720-0.101894 i$ & $0.036313-0.101387 i$ & $0.600\%$\\
$0.1$ & $0.065132-0.178166 i$ & $0.064713-0.177577 i$ & $0.381\%$\\
$0.15$ & $0.088376-0.237482 i$ & $0.088112-0.237051 i$ & $0.200\%$\\
$0.2$ & $0.108154-0.284908 i$ & $0.108044-0.284689 i$ & $0.0803\%$\\
$0.25$ & $0.125517-0.323589 i$ & $0.125498-0.323534 i$ & $0.0167\%$\\
$0.3$ & $0.141173-0.355577 i$ & $0.141173-0.355577 i$ & $0.0001\%$\\
$0.35$ & $0.155593-0.382185 i$ & $0.155596-0.382159 i$ & $0.0063\%$\\
$0.4$ & $0.169338-0.403619 i$ & $0.169194-0.404259 i$ & $0.150\%$\\
$0.45$ & $0.180848-0.423416 i$ & $0.182388-0.422598 i$ & $0.379\%$\\
$0.5$ & $0.194231-0.437667 i$ & $0.195270-0.437663 i$ & $0.217\%$\\
$0.55$ & $0.206869-0.448363 i$ & $0.206740-0.449018 i$ & $0.135\%$\\
$0.6$ & $0.220036-0.455609 i$ & $0.219984-0.455604 i$ & $0.0102\%$\\
$0.65$ & $0.233415-0.459406 i$ & $0.233398-0.459386 i$ & $0.00513\%$\\
$0.7$ & $0.246193-0.459071 i$ & $0.246433-0.458731 i$ & $0.0798\%$\\
$0.75$ & $0.257740-0.452108 i$ & $0.258049-0.452311 i$ & $0.0712\%$\\
$0.8$ & $0.264546-0.438948 i$ & $0.264674-0.439066 i$ & $0.0339\%$\\
$0.85$ & $0.262073-0.425186 i$ & $0.261546-0.425756 i$ & $0.155\%$\\
\hline
\end{tabular}
\caption{Overtone QNMs of the $\ell=2$, $n=1,2,3$ perturbations of the Jusufi--Singleton black hole ($M=1$), calculated using the WKB formula at different orders and Pad\'e approximants. The deviation is given in percent.}
\end{table}


\section{Excitation factors}\label{sec:excitation}

The quasinormal frequencies determine the pole positions of the Green function, but not by themselves the strength with which a given pole contributes to a ringdown signal.  For that purpose, one needs the excitation factor, namely the residue of the frequency-domain Green function at the quasinormal pole.  In this section, we compute this quantity for the fundamental axial gravitational modes with $\ell=2,3,4$.

Let $\Psi^{\rm in}$ denote the solution which is purely ingoing at the event horizon,
\begin{equation}
 \Psi^{\rm in}\sim e^{-\imo\omega r_*},\qquad r\to r_+ .
\end{equation}
At a large radius, the same solution is decomposed as
\begin{equation}\label{Apmdef}
\Psi^{\rm in}\sim A^{(-)}(\omega)e^{-\imo\omega r_*}
+A^{(+)}(\omega)e^{+\imo\omega r_*},\qquad r\to\infty .
\end{equation}
The quasinormal modes are the zeros of the incoming amplitude, $A^{(-)}(\omega_{\ell n})=0$, and the corresponding excitation factor is
\begin{equation}\label{Bfactor}
B_{\ell n}=\left.
\frac{A^{(+)}(\omega)}{2\omega\,dA^{(-)}(\omega)/d\omega}
\right|_{\omega=\omega_{\ell n}} .
\end{equation}
Equivalently, after introducing the ADM-scaled frequency $\Omega=M_{\rm ADM}\omega$, one may write the denominator as $2\Omega\,dA^{(-)}/d\Omega$; the value of $B_{\ell n}$ is unchanged by this change of frequency variable.  The residue is dimensionless, but it is not independent of the additive constant in the tortoise coordinate.  To make the normalization definite, we use ADM variables
\begin{equation}\label{ADM-tortoise-convention}
\begin{gathered}
 \bar r=\frac{r}{M_{\rm ADM}},\qquad
 \bar r_* =\frac{r_*}{M_{\rm ADM}},\\
 \bar r_* -\bar r-2\ln \bar r \longrightarrow0,
 \qquad r\to\infty .
\end{gathered}
\end{equation}
A different additive convention for $r_*$ would multiply all tabulated residues by a known complex factor.  The products with source integrals or initial-data coefficients are invariant, so the excitation factors below should be compared using the same convention.

The numerical implementation follows directly from the axial gravitational equation.  We factor out the horizon phase,
\begin{equation}
 \Psi^{\rm in}=e^{-\imo\omega r_*}H(r),
\end{equation}
which gives
\begin{equation}\label{Heq}
 f^2H''+f\left(f'-2\imo\omega\right)H'-V_\ell(r)H=0 .
\end{equation}
The regular expansion at the outer horizon starts as
\begin{equation}
\begin{split}
H(r)=1&+
\frac{\ell(\ell+1)-2-r_+f'(r_+)}
 {r_+^2\left[f'(r_+)-2\imo\omega\right]}
 (r-r_+)\\
&+\Order{(r-r_+)^2} .
\end{split}
\end{equation}
After integrating Eq.~(\ref{Heq}) outward, the solution is matched to two asymptotic branches of the form
\begin{eqnarray}
\Psi^{\rm in}&=&A^{(-)}e^{-\imo\omega r_*}
\left(1+\sum_{k=1}^{K}\frac{a_k^{(-)}}{r^k}\right)\nonumber\\
&&+A^{(+)}e^{+\imo\omega r_*}
\left(1+\sum_{k=1}^{K}\frac{a_k^{(+)}}{r^k}\right) .
\end{eqnarray}
The coefficients are obtained recursively from the large-$r$ expansions of $f(r)$ and $V_\ell(r)$.  We used $K=12$ and refined the WKB--Pad\'e frequencies of Sec.~\ref{sec:qnm} by solving $A^{(-)}=0$ before evaluating the residue with a five-point finite-difference derivative.  The matching radii satisfy $R/M_{\rm ADM}\gtrsim30$, and the largest residual $|A^{(-)}|$ at the roots in Table~\ref{tab:grav-excitation-adm} is below $4\times10^{-10}$.

\begin{table}
\centering
\scriptsize
\setlength{\tabcolsep}{3.5pt}
\renewcommand{\arraystretch}{0.88}
\begin{tabular}{c c c c}
\hline
$l_0/M_{\rm ADM}$ & $M_{\rm ADM}\omega_{\ell0}$ & $B_{\ell0}$ & $|B_{\ell0}|$\\
\hline
\multicolumn{4}{c}{$\ell=2$}\\
\hline
$0.00726$ & $0.37505-0.08907\imo$ & $0.06446-0.07707\imo$ & $0.10048$\\
$0.02535$ & $0.37795-0.08930\imo$ & $0.06504-0.07672\imo$ & $0.10058$\\
$0.05062$ & $0.38144-0.08954\imo$ & $0.06579-0.07625\imo$ & $0.10071$\\
$0.08089$ & $0.38517-0.08976\imo$ & $0.06665-0.07566\imo$ & $0.10083$\\
$0.11478$ & $0.38901-0.08993\imo$ & $0.06762-0.07495\imo$ & $0.10095$\\
$0.15138$ & $0.39293-0.09005\imo$ & $0.06872-0.07409\imo$ & $0.10105$\\
$0.19006$ & $0.39695-0.09008\imo$ & $0.06995-0.07303\imo$ & $0.10113$\\
$0.23037$ & $0.40115-0.09003\imo$ & $0.07136-0.07172\imo$ & $0.10118$\\
$0.27199$ & $0.40558-0.08986\imo$ & $0.07296-0.07009\imo$ & $0.10117$\\
$0.31465$ & $0.41034-0.08956\imo$ & $0.07478-0.06804\imo$ & $0.10110$\\
$0.35819$ & $0.41552-0.08908\imo$ & $0.07683-0.06545\imo$ & $0.10093$\\
$0.40245$ & $0.42127-0.08839\imo$ & $0.07912-0.06212\imo$ & $0.10060$\\
$0.44732$ & $0.42772-0.08740\imo$ & $0.08161-0.05782\imo$ & $0.10001$\\
$0.49270$ & $0.43509-0.08600\imo$ & $0.08416-0.05215\imo$ & $0.09901$\\
$0.53852$ & $0.44364-0.08396\imo$ & $0.08644-0.04459\imo$ & $0.09727$\\
$0.58473$ & $0.45377-0.08089\imo$ & $0.08762-0.03437\imo$ & $0.09412$\\
$0.60331$ & $0.45836-0.07921\imo$ & $0.08738-0.02937\imo$ & $0.09218$\\
\hline
\multicolumn{4}{c}{$\ell=3$}\\
\hline
$0.00726$ & $0.60161-0.09282\imo$ & $-0.03161+0.07567\imo$ & $0.08201$\\
$0.02535$ & $0.60617-0.09305\imo$ & $-0.03235+0.07544\imo$ & $0.08208$\\
$0.05062$ & $0.61166-0.09330\imo$ & $-0.03332+0.07511\imo$ & $0.08217$\\
$0.08089$ & $0.61752-0.09353\imo$ & $-0.03446+0.07469\imo$ & $0.08226$\\
$0.11478$ & $0.62353-0.09371\imo$ & $-0.03576+0.07417\imo$ & $0.08234$\\
$0.15138$ & $0.62966-0.09383\imo$ & $-0.03726+0.07350\imo$ & $0.08240$\\
$0.19006$ & $0.63595-0.09387\imo$ & $-0.03898+0.07265\imo$ & $0.08245$\\
$0.23037$ & $0.64249-0.09381\imo$ & $-0.04097+0.07156\imo$ & $0.08246$\\
$0.27199$ & $0.64938-0.09364\imo$ & $-0.04329+0.07016\imo$ & $0.08244$\\
$0.31465$ & $0.65677-0.09334\imo$ & $-0.04598+0.06833\imo$ & $0.08236$\\
$0.35819$ & $0.66481-0.09286\imo$ & $-0.04909+0.06592\imo$ & $0.08219$\\
$0.40245$ & $0.67370-0.09216\imo$ & $-0.05266+0.06271\imo$ & $0.08189$\\
$0.44732$ & $0.68369-0.09115\imo$ & $-0.05669+0.05838\imo$ & $0.08138$\\
$0.49270$ & $0.69511-0.08973\imo$ & $-0.06109+0.05246\imo$ & $0.08053$\\
$0.53852$ & $0.70841-0.08766\imo$ & $-0.06557+0.04421\imo$ & $0.07908$\\
$0.58473$ & $0.72427-0.08456\imo$ & $-0.06924+0.03255\imo$ & $0.07651$\\
$0.60331$ & $0.73155-0.08285\imo$ & $-0.07004+0.02662\imo$ & $0.07493$\\
\hline
\multicolumn{4}{c}{$\ell=4$}\\
\hline
$0.00726$ & $0.81208-0.09428\imo$ & $0.00833-0.07066\imo$ & $0.07115$\\
$0.02535$ & $0.81819-0.09451\imo$ & $0.00919-0.07062\imo$ & $0.07121$\\
$0.05062$ & $0.82554-0.09476\imo$ & $0.01033-0.07053\imo$ & $0.07128$\\
$0.08089$ & $0.83337-0.09498\imo$ & $0.01168-0.07039\imo$ & $0.07135$\\
$0.11478$ & $0.84141-0.09516\imo$ & $0.01324-0.07018\imo$ & $0.07141$\\
$0.15138$ & $0.84960-0.09527\imo$ & $0.01505-0.06986\imo$ & $0.07146$\\
$0.19006$ & $0.85800-0.09531\imo$ & $0.01716-0.06940\imo$ & $0.07149$\\
$0.23037$ & $0.86672-0.09525\imo$ & $0.01964-0.06875\imo$ & $0.07150$\\
$0.27199$ & $0.87591-0.09507\imo$ & $0.02254-0.06782\imo$ & $0.07147$\\
$0.31465$ & $0.88574-0.09475\imo$ & $0.02597-0.06650\imo$ & $0.07139$\\
$0.35819$ & $0.89645-0.09426\imo$ & $0.02999-0.06461\imo$ & $0.07123$\\
$0.40245$ & $0.90828-0.09354\imo$ & $0.03468-0.06190\imo$ & $0.07095$\\
$0.44732$ & $0.92156-0.09252\imo$ & $0.04008-0.05799\imo$ & $0.07049$\\
$0.49270$ & $0.93674-0.09107\imo$ & $0.04614-0.05229\imo$ & $0.06974$\\
$0.53852$ & $0.95443-0.08897\imo$ & $0.05256-0.04389\imo$ & $0.06847$\\
$0.58473$ & $0.97557-0.08583\imo$ & $0.05833-0.03138\imo$ & $0.06623$\\
$0.60331$ & $0.98531-0.08410\imo$ & $0.05992-0.02482\imo$ & $0.06486$\\
\hline
\end{tabular}
\caption{Fundamental axial gravitational excitation factors in ADM units.  The first column uses the dimensionless zero-point length $l_0/M_{\rm ADM}$, the frequency column gives the refined value of $M_{\rm ADM}\omega_{\ell0}$, and the residues use the ADM tortoise convention of Eq.~(\ref{ADM-tortoise-convention}). The largest zero-point length shown is $0.82$.}
\label{tab:grav-excitation-adm}
\end{table}

The first entry in each multipole block also provides a check of the Schwarzschild limit.  At $l_0/M_{\rm ADM}=0.00726$, the frequency column gives $M_{\rm ADM}\omega_{20}=0.37505-0.08907\imo$, $M_{\rm ADM}\omega_{30}=0.60161-0.09282\imo$, and $M_{\rm ADM}\omega_{40}=0.81208-0.09428\imo$.  These values are already close to the standard Schwarzschild axial gravitational frequencies, $0.37367-0.08896\imo$, $0.59944-0.09270\imo$, and $0.80918-0.09416\imo$, respectively~\cite{Berti:2009kk}.  The complex relative differences are about $0.36\%$ for all three multipoles, which is consistent with the fact that the table uses a small but finite value of $l_0$ rather than the exact $l_0\to0$ limit. In the exact $l_0 = 0$ limit, we recover the Schwarzschild values of excitation factors. 

Several features are apparent from Table~\ref{tab:grav-excitation-adm}.  First, the ADM-scaled frequencies follow the same trend already seen in Sec.~\ref{sec:qnm}: the real part grows monotonically with $l_0/M_{\rm ADM}$, whereas the damping rate has only a shallow maximum before decreasing near the upper end of the interval.  The residue is less sensitive to the zero-point length than the frequency, although its modulus decreases more clearly in the near-extremal regime.  Over the full range now shown, $|B_{20}|$ changes from about $0.100$ to $0.092$, $|B_{30}|$ from about $0.082$ to $0.075$, and $|B_{40}|$ from about $0.071$ to $0.065$.  Thus, the self-energy deformation still shifts the pole positions more visibly than it changes the intrinsic residue of the fundamental gravitational pole.

The ordering with multipole number is also robust: $|B_{20}|>|B_{30}|>|B_{40}|$ for every value of $l_0/M_{\rm ADM}$ in Table~\ref{tab:grav-excitation-adm}.  In this Green-function normalization, the quadrupole fundamental mode is therefore the most strongly excited of the three axial gravitational modes considered here, while the higher multipoles have smaller residues.  The complex phases vary smoothly with the deformation parameter.  This phase motion should not be interpreted as a separate stability criterion; it reflects the scattering normalization and the convention for $r_*$.  A physical waveform amplitude is obtained only after multiplying $B_{\ell0}$ by the source or initial-data integral appropriate to the perturbation.

\section{Conclusions}\label{sec:conclusions}

We have analyzed axial gravitational perturbations of the regular self-energy black hole generated by a zero-point length $l_0$.  The calculation is naturally organized in terms of the ADM mass, because the bare parameter $M$ is not the mass measured at infinity: for $M=1$ one has $M_{\rm ADM}=1+3\pi/(32l_0)$.  Expressing the spectra in the dimensionless variables $M_{\rm ADM}\omega$ and $l_0/M_{\rm ADM}$ therefore separates the physical change of the exterior potential from the trivial rescaling of the total mass.  In this normalization, the Schwarzschild Regge-Wheeler problem is recovered as $l_0\to0$ at fixed ADM mass, while finite $l_0$ provides a controlled deformation of the gravitational barrier.

The WKB--Pad\'e spectrum exhibits a clear dependence on the zero-point length.  For the fundamental modes with $\ell=2,3,4$, $M_{\rm ADM}{\rm Re}(\omega)$ increases monotonically up to $l_0=0.85$, while the damping rate has a shallow maximum and then decreases toward extremality.  For the $\ell=2$ overtones, the real parts of the $n=1$ and $n=2$ modes remain monotonic, whereas the $n=3$ mode turns over between $l_0=0.80$ and $0.85$.  The close agreement between the two Pad\'e-resummed WKB orders supports the numerical stability of these trends.

The time-domain evolutions independently confirm the two fundamental frequencies considered here, with complex relative differences of $1.20\times10^{-3}\%$ and $4.91\times10^{-3}\%$ from the WKB--Pad\'e values.  Their late-time behavior follows the Price-law exponent $t^{-(2\ell+3)}$, showing that the regular core does not change the leading asymptotic decay.  The excitation factors vary more weakly than the pole positions: over the range in Table~\ref{tab:grav-excitation-adm}, their magnitudes decrease from approximately $0.100$ to $0.092$ for $\ell=2$, from $0.082$ to $0.075$ for $\ell=3$, and from $0.071$ to $0.065$ for $\ell=4$, while the ordering $|B_{20}|>|B_{30}|>|B_{40}|$ is preserved.

These results provide the ingredients for future tests based on the deformation parameter $l_0/M_{\rm ADM}$.  A Bayesian ringdown analysis could use the tabulated frequencies, with the excitation factors serving only as guidance for amplitude priors because physical amplitudes also depend on the source and initial data.  An independent verification of the overtones can be obtained by Taylor-expanding the metric function in a compact radial coordinate, casting the wave equation into a form with rational coefficients, and applying Leaver's continued-fraction method~\cite{Leaver:1986gd}, which was used in a great number of studies for finding precise quasinormal modes \cite{Nollert:1992ifk,Dias:2021yju,Onozawa:1996ux,Bolokhov:2023bwm,Bolokhov:2023dxq,Zinhailo:2024kbq,Kanti:2006ua,Konoplya:2004uk}.  Further extensions include a direct test of the QNM--grey-body-factor correspondence~\cite{Lutfuoglu:2025mqa,Han:2025cal,Dubinsky:2024vbn,Konoplya:2024lir,Konoplya:2024vuj,Han:2026fpn,Malik:2024cgb,Skvortsova:2024msa,Bolokhov:2024otn}, perturbations of massive fields~\cite{Ohashi:2004wr,Koyama:2001qw,Konoplya:2007zx,Moderski:2001tk,Rogatko:2007zz,Konoplya:2017tvu,Lutfuoglu:2026uzy,Lutfuoglu:2026gis}, and the rotating generalization required for precision comparisons with astrophysical remnants.

After the first version of this work appeared on arXiv, several further studies investigating the spectrum and scattering properties of this spacetime were published. These include analyses of perturbations and spectra of test fields \cite{Skvortsova:2026ryl}, quasinormal modes, scattering, and quasi-bound states of massive fields \cite{Dubinsky:2026nww}, as well as Hawking radiation \cite{Lutfuoglu:2026asf}.

\section*{Declaration of Competing Interest}
The authors declare that they have no known competing financial interests or personal relationships that could have appeared to influence the work reported in this paper.

\section*{Data Availability}
No data was used for the research described in the article.

\begin{acknowledgments}
B. C. L. is grateful to the Excellence project FoS UHK 2205/2025-2026 for the financial support.
\end{acknowledgments}

\bibliography{bibliographyGRAV}

@article{Berti:2009kk,
    author = "Berti, Emanuele and Cardoso, Vitor and Starinets, Andrei O.",
    title = "{Quasinormal modes of black holes and black branes}",
    eprint = "0905.2975",
    archivePrefix = "arXiv",
    primaryClass = "gr-qc",
    doi = "10.1088/0264-9381/26/16/163001",
    journal = "Class. Quant. Grav.",
    volume = "26",
    pages = "163001",
    year = "2009"
}

@article{Konoplya:2024lch,
    author = "Konoplya, R. A. and Stashko, O. S.",
    title = "{Probing the effective quantum gravity via quasinormal modes and shadows of black holes}",
    eprint = "2408.02578",
    archivePrefix = "arXiv",
    primaryClass = "gr-qc",
    doi = "10.1103/PhysRevD.111.104055",
    journal = "Phys. Rev. D",
    volume = "111",
    number = "10",
    pages = "104055",
    year = "2025"
}

@article{Konoplya:2011qq,
    author = "Konoplya, R. A. and Zhidenko, A.",
    title = "{Quasinormal modes of black holes: From astrophysics to string theory}",
    eprint = "1102.4014",
    archivePrefix = "arXiv",
    primaryClass = "gr-qc",
    doi = "10.1103/RevModPhys.83.793",
    journal = "Rev. Mod. Phys.",
    volume = "83",
    pages = "793--836",
    year = "2011"
}

@article{Hayward:2005gi,
    author = "Hayward, Sean A.",
    title = "{Formation and evaporation of regular black holes}",
    eprint = "gr-qc/0506126",
    archivePrefix = "arXiv",
    doi = "10.1103/PhysRevLett.96.031103",
    journal = "Phys. Rev. Lett.",
    volume = "96",
    pages = "031103",
    year = "2006"
}

@article{Lin:2013ofa,
    author = "Lin, Kai and Li, Jin and Yang, Shuzheng",
    title = "{Quasinormal Modes of Hayward Regular Black Hole}",
    doi = "10.1007/s10773-013-1682-4",
    journal = "Int. J. Theor. Phys.",
    volume = "52",
    pages = "3771--3778",
    year = "2013"
}

@article{Macedo:2016yyo,
    author = "Macedo, Caio F. B. and Crispino, Lu{\'\i}s C. B. and de Oliveira, Ednilton S.",
    editor = "Herdeiro, Carlos A. R. and Berti, Emanuele and Cardoso, Vitor and Crispino, Luis C. B. and Gualtieri, Leonardo and Sperhake, Ulrich",
    title = "{Scalar waves in regular Bardeen black holes: Scattering, absorption and quasinormal modes}",
    eprint = "1605.00123",
    archivePrefix = "arXiv",
    primaryClass = "gr-qc",
    doi = "10.1142/S021827181641008X",
    journal = "Int. J. Mod. Phys. D",
    volume = "25",
    number = "09",
    pages = "1641008",
    year = "2016"
}

@article{Konoplya:2023aph,
    author = "Konoplya, R. A. and Stuchlik, Z. and Zhidenko, A. and Zinhailo, A. F.",
    title = "{Quasinormal modes of renormalization group improved Dymnikova regular black holes}",
    eprint = "2303.01987",
    archivePrefix = "arXiv",
    primaryClass = "gr-qc",
    doi = "10.1103/PhysRevD.107.104050",
    journal = "Phys. Rev. D",
    volume = "107",
    number = "10",
    pages = "104050",
    year = "2023"
}

@article{Konoplya:2023ppx,
    author = "Konoplya, R. A.",
    title = "{Quasinormal modes and grey-body factors of regular black holes with a scalar hair from the Effective Field Theory}",
    eprint = "2305.09187",
    archivePrefix = "arXiv",
    primaryClass = "gr-qc",
    doi = "10.1088/1475-7516/2023/07/001",
    journal = "JCAP",
    volume = "07",
    pages = "001",
    year = "2023"
}

@article{Konoplya:2019hlu,
    author = "Konoplya, R. A. and Zhidenko, A. and Zinhailo, A. F.",
    title = "{Higher order WKB formula for quasinormal modes and grey-body factors: recipes for quick and accurate calculations}",
    eprint = "1904.10333",
    archivePrefix = "arXiv",
    primaryClass = "gr-qc",
    doi = "10.1088/1361-6382/ab2e25",
    journal = "Class. Quant. Grav.",
    volume = "36",
    pages = "155002",
    year = "2019"
}

@article{Iyer:1986np,
    author = "Iyer, Sai and Will, Clifford M.",
    title = "{Black Hole Normal Modes: A {WKB} Approach. 1. Foundations and Application of a Higher Order {WKB} Analysis of Potential Barrier Scattering}",
    reportNumber = "Print-86-1482 (WASH. U., ST. LOUIS)",
    doi = "10.1103/PhysRevD.35.3621",
    journal = "Phys. Rev. D",
    volume = "35",
    pages = "3621",
    year = "1987"
}

@article{Konoplya:2003ii,
    author = "Konoplya, R. A.",
    title = "{Quasinormal behavior of the d-dimensional Schwarzschild black hole and higher order WKB approach}",
    eprint = "gr-qc/0303052",
    archivePrefix = "arXiv",
    doi = "10.1103/PhysRevD.68.024018",
    journal = "Phys. Rev. D",
    volume = "68",
    pages = "024018",
    year = "2003"
}

@article{Matyjasek:2017psv,
    author = "Matyjasek, Jerzy and Opala, Micha{\l}",
    title = "{Quasinormal modes of black holes. The improved semianalytic approach}",
    eprint = "1704.00361",
    archivePrefix = "arXiv",
    primaryClass = "gr-qc",
    doi = "10.1103/PhysRevD.96.024011",
    journal = "Phys. Rev. D",
    volume = "96",
    number = "2",
    pages = "024011",
    year = "2017"
}

@article{Kokkotas:2010zd,
    author = "Kokkotas, K. D. and Konoplya, R. A. and Zhidenko, A.",
    title = "{Quasinormal modes, scattering and Hawking radiation of Kerr-Newman black holes in a magnetic field}",
    eprint = "1011.1843",
    archivePrefix = "arXiv",
    primaryClass = "gr-qc",
    doi = "10.1103/PhysRevD.83.024031",
    journal = "Phys. Rev. D",
    volume = "83",
    pages = "024031",
    year = "2011"
}

@article{Konoplya:2010vz,
    author = "Konoplya, R. A. and Zhidenko, A.",
    title = "{Long life of Gauss-Bonnet corrected black holes}",
    eprint = "1004.3772",
    archivePrefix = "arXiv",
    primaryClass = "hep-th",
    doi = "10.1103/PhysRevD.82.084003",
    journal = "Phys. Rev. D",
    volume = "82",
    pages = "084003",
    year = "2010"
}

@article{Guo:2020blq,
    author = "Guo, Hong and Liu, Hang and Kuang, Xiao-Mei and Wang, Bin",
    title = "{Acoustic black hole in Schwarzschild spacetime: quasi-normal modes, analogous Hawking radiation and shadows}",
    eprint = "2007.04197",
    archivePrefix = "arXiv",
    primaryClass = "gr-qc",
    doi = "10.1103/PhysRevD.102.124019",
    journal = "Phys. Rev. D",
    volume = "102",
    pages = "124019",
    year = "2020"
}

@article{Tan:2022vfe,
    author = "Tan, Qin and Guo, Wen-Di and Liu, Yu-Xiao",
    title = "{Sound from extra dimensions: Quasinormal modes of a thick brane}",
    eprint = "2205.05255",
    archivePrefix = "arXiv",
    primaryClass = "gr-qc",
    doi = "10.1103/PhysRevD.106.044038",
    journal = "Phys. Rev. D",
    volume = "106",
    number = "4",
    pages = "044038",
    year = "2022"
}

@article{Albuquerque:2023lhm,
    author = "Albuquerque, Saulo and Lobo, Iarley P. and Bezerra, Valdir B.",
    title = "{Massless Dirac perturbations in a consistent model of loop quantum gravity black hole: quasinormal modes and particle emission rates}",
    eprint = "2301.07746",
    archivePrefix = "arXiv",
    primaryClass = "gr-qc",
    doi = "10.1088/1361-6382/ace7a8",
    journal = "Class. Quant. Grav.",
    volume = "40",
    number = "17",
    pages = "174001",
    year = "2023"
}

@article{Fernando:2016ftj,
    author = "Fernando, Sharmanthie",
    title = "{Quasinormal modes of dilaton-de Sitter black holes: scalar perturbations}",
    eprint = "1601.06407",
    archivePrefix = "arXiv",
    primaryClass = "gr-qc",
    doi = "10.1007/s10714-016-2020-y",
    journal = "Gen. Rel. Grav.",
    volume = "48",
    number = "3",
    pages = "24",
    year = "2016"
}

@article{Gundlach:1993tp,
    author = "Gundlach, Carsten and Price, Richard H. and Pullin, Jorge",
    title = "{Late time behavior of stellar collapse and explosions: 1. Linearized perturbations}",
    eprint = "gr-qc/9307009",
    archivePrefix = "arXiv",
    reportNumber = "NSF-ITP-93-84",
    doi = "10.1103/PhysRevD.49.883",
    journal = "Phys. Rev. D",
    volume = "49",
    pages = "883--889",
    year = "1994"
}

@article{Konoplya:2014lha,
    author = "Konoplya, R. A. and Zhidenko, A.",
    title = "{Charged scalar field instability between the event and cosmological horizons}",
    eprint = "1406.0019",
    archivePrefix = "arXiv",
    primaryClass = "hep-th",
    doi = "10.1103/PhysRevD.90.064048",
    journal = "Phys. Rev. D",
    volume = "90",
    number = "6",
    pages = "064048",
    year = "2014"
}

@article{Konoplya:2005et,
    author = "Konoplya, R. A. and Molina, C.",
    title = "{The Ringing wormholes}",
    eprint = "gr-qc/0504139",
    archivePrefix = "arXiv",
    doi = "10.1103/PhysRevD.71.124009",
    journal = "Phys. Rev. D",
    volume = "71",
    pages = "124009",
    year = "2005"
}

@article{Konoplya:2018yrp,
    author = "Konoplya, R. A. and Stuchl{\'\i}k, Z. and Zhidenko, A.",
    title = "{Echoes of compact objects: new physics near the surface and matter at a distance}",
    eprint = "1810.01295",
    archivePrefix = "arXiv",
    primaryClass = "gr-qc",
    doi = "10.1103/PhysRevD.99.024007",
    journal = "Phys. Rev. D",
    volume = "99",
    number = "2",
    pages = "024007",
    year = "2019"
}

@article{Abdalla:2012si,
    author = "Abdalla, Elcio and Piedra, Owen Pavel Fernandez and Nu{\~n}ez, Fidel Sosa and de Oliveira, Jeferson",
    title = "{Scalar field propagation in higher dimensional black holes at a Lifshitz point}",
    eprint = "1211.3390",
    archivePrefix = "arXiv",
    primaryClass = "gr-qc",
    reportNumber = "GEA-UCF-2012-03",
    doi = "10.1103/PhysRevD.88.064035",
    journal = "Phys. Rev. D",
    volume = "88",
    number = "6",
    pages = "064035",
    year = "2013"
}

@article{Aneesh:2018hlp,
    author = "Aneesh, S. and Bose, Sukanta and Kar, Sayan",
    title = "{Gravitational waves from quasinormal modes of a class of Lorentzian wormholes}",
    eprint = "1803.10204",
    archivePrefix = "arXiv",
    primaryClass = "gr-qc",
    doi = "10.1103/PhysRevD.97.124004",
    journal = "Phys. Rev. D",
    volume = "97",
    number = "12",
    pages = "124004",
    year = "2018"
}

@article{Momennia:2022tug,
    author = "Momennia, Mehrab",
    title = "{Quasinormal modes of self-dual black holes in loop quantum gravity}",
    eprint = "2204.03259",
    archivePrefix = "arXiv",
    primaryClass = "gr-qc",
    doi = "10.1103/PhysRevD.106.024052",
    journal = "Phys. Rev. D",
    volume = "106",
    number = "2",
    pages = "024052",
    year = "2022"
}

@article{Leaver:1986gd,
    author = "Leaver, Edward W.",
    title = "{Spectral decomposition of the perturbation response of the Schwarzschild geometry}",
    doi = "10.1103/PhysRevD.34.384",
    journal = "Phys. Rev. D",
    volume = "34",
    pages = "384--408",
    year = "1986"
}

@article{Jusufi:2025selfenergy,
    author = "Jusufi, Kimet and Singleton, Douglas",
    title = "{Regular black holes with gravitational self-energy as dark matter}",
    eprint = "2509.13335",
    archivePrefix = "arXiv",
    primaryClass = "gr-qc",
    doi = "10.1140/epjc/s10052-026-15787-w",
    journal = "Eur. Phys. J. C",
    volume = "86",
    number = "6",
    pages = "530",
    year = "2026"
}

@article{Padmanabhan:1997zpl,
    author = "Padmanabhan, T.",
    title = "{Duality and zero-point length of spacetime}",
    eprint = "hep-th/9608182",
    archivePrefix = "arXiv",
    doi = "10.1103/PhysRevLett.78.1854",
    journal = "Phys. Rev. Lett.",
    volume = "78",
    pages = "1854--1857",
    year = "1997"
}

@article{Nicolini:2022zpl,
    author = "Nicolini, Piero",
    title = "{Quantum gravity and the zero point length}",
    eprint = "2208.05390",
    archivePrefix = "arXiv",
    primaryClass = "hep-th",
    doi = "10.1007/s10714-022-02995-4",
    journal = "Gen. Rel. Grav.",
    volume = "54",
    number = "9",
    pages = "106",
    year = "2022"
}

@article{Nicolini:2019tduality,
    author = "Nicolini, Piero and Spallucci, Euro and Wondrak, Michael F.",
    title = "{Quantum corrected black holes from string T-duality}",
    eprint = "1902.11242",
    archivePrefix = "arXiv",
    primaryClass = "gr-qc",
    doi = "10.1016/j.physletb.2019.134888",
    journal = "Phys. Lett. B",
    volume = "797",
    pages = "134888",
    year = "2019"
}

@article{Gaete:2022tduality,
    author = "Gaete, Patricio and Jusufi, Kimet and Nicolini, Piero",
    title = "{Charged black holes from T-duality}",
    eprint = "2205.15441",
    archivePrefix = "arXiv",
    primaryClass = "hep-th",
    doi = "10.1016/j.physletb.2022.137546",
    journal = "Phys. Lett. B",
    volume = "835",
    pages = "137546",
    year = "2022"
}

@article{Hehl:2009nonlocal,
    author = "Hehl, Friedrich W. and Mashhoon, Bahram",
    title = "{Nonlocal gravity simulates dark matter}",
    eprint = "0812.1059",
    archivePrefix = "arXiv",
    primaryClass = "gr-qc",
    doi = "10.1016/j.physletb.2009.02.033",
    journal = "Phys. Lett. B",
    volume = "673",
    pages = "279--282",
    year = "2009"
}

@article{JusufiNicolini:2025geodesic,
    author = "Jusufi, Kimet and Nicolini, Piero",
    title = "{Geodesic completeness from string T-duality}",
    eprint = "2410.19613",
    archivePrefix = "arXiv",
    primaryClass = "hep-th",
    doi = "10.1140/epjc/s10052-025-15018-8",
    journal = "Eur. Phys. J. C",
    volume = "85",
    number = "11",
    pages = "1291",
    year = "2025"
}

@misc{Jusufi:2022three,
    author = "Jusufi, Kimet",
    title = "{Regular black holes in three dimensions and the zero point length}",
    eprint = "2209.04433",
    archivePrefix = "arXiv",
    primaryClass = "gr-qc",
    year = "2022"
}

@article{Jusufi:2023collapse,
    author = "Jusufi, Kimet",
    title = "{Avoidance of singularity during the gravitational collapse with string T-duality effects}",
    eprint = "2301.03590",
    archivePrefix = "arXiv",
    primaryClass = "gr-qc",
    doi = "10.3390/universe9010041",
    journal = "Universe",
    volume = "9",
    number = "1",
    pages = "41",
    year = "2023"
}

@article{JusufiSingletonLobo:2026collapse,
    author = "Jusufi, Kimet and Singleton, Douglas and Lobo, Francisco S. N.",
    title = "{Spontaneous wave function collapse from non-local gravitational self-energy}",
    eprint = "2512.15393",
    archivePrefix = "arXiv",
    primaryClass = "gr-qc",
    doi = "10.1016/j.physletb.2026.140475",
    journal = "Phys. Lett. B",
    volume = "877",
    pages = "140475",
    year = "2026"
}

@article{Ashtekar:2018lag,
    author = "Ashtekar, Abhay and Olmedo, Javier and Singh, Parampreet",
    title = "{Quantum Extension of the Kruskal Space-time}",
    eprint = "1806.02406",
    archivePrefix = "arXiv",
    primaryClass = "gr-qc",
    doi = "10.1103/PhysRevD.98.126003",
    journal = "Phys. Rev. D",
    volume = "98",
    number = "12",
    pages = "126003",
    year = "2018"
}

@article{Ashtekar:2018cay,
    author = "Ashtekar, Abhay and Olmedo, Javier and Singh, Parampreet",
    title = "{Quantum Transfiguration of Kruskal Black Holes}",
    eprint = "1806.00648",
    archivePrefix = "arXiv",
    primaryClass = "gr-qc",
    doi = "10.1103/PhysRevLett.121.241301",
    journal = "Phys. Rev. Lett.",
    volume = "121",
    number = "24",
    pages = "241301",
    year = "2018"
}

@article{Chakraborty:2024gcr,
    author = "Chakraborty, Sumanta and Comp{\`e}re, Geoffrey and Machet, Ludovico",
    title = "{Tidal Love numbers and quasinormal modes of the Schwarzschild-Hernquist black hole}",
    eprint = "2412.14831",
    archivePrefix = "arXiv",
    primaryClass = "gr-qc",
    doi = "10.1103/4p2c-rwdh",
    journal = "Phys. Rev. D",
    volume = "112",
    number = "2",
    pages = "024015",
    year = "2025"
}

@article{Bolokhov:2025lnt,
    author = "Bolokhov, S. V. and Skvortsova, Milena",
    title = "{Gravitational Quasinormal Modes and Grey-Body Factors of Bonanno--Reuter Regular Black Holes}",
    eprint = "2507.07196",
    archivePrefix = "arXiv",
    primaryClass = "gr-qc",
    doi = "10.53941/ijgtp.2025.100003",
    journal = "Int. J. Grav. Theor. Phys.",
    volume = "1",
    number = "1",
    pages = "3",
    year = "2025"
}

@misc{Zhao:2026eti,
    author = "Zhao, Yu-Qian and Pani, Paolo",
    title = "{Quasinormal modes and tidal responses of black holes in generic anisotropic matter environments}",
    eprint = "2606.11380",
    archivePrefix = "arXiv",
    primaryClass = "gr-qc",
    month = "6",
    year = "2026"
}

@article{Regge:1957td,
    author = "Regge, Tullio and Wheeler, John A.",
    title = "{Stability of a Schwarzschild singularity}",
    doi = "10.1103/PhysRev.108.1063",
    journal = "Phys. Rev.",
    volume = "108",
    pages = "1063--1069",
    year = "1957"
}

@article{Bronnikov:2012ch,
    author = "Bronnikov, K. A. and Konoplya, R. A. and Zhidenko, A.",
    title = "{Instabilities of wormholes and regular black holes supported by a phantom scalar field}",
    eprint = "1205.2224",
    archivePrefix = "arXiv",
    primaryClass = "gr-qc",
    doi = "10.1103/PhysRevD.86.024028",
    journal = "Phys. Rev. D",
    volume = "86",
    pages = "024028",
    year = "2012"
}

@article{Chen:2019iuo,
    author = "Chen, Che-Yu and Chen, Pisin",
    title = "{Gravitational perturbations of nonsingular black holes in conformal gravity}",
    eprint = "1902.01678",
    archivePrefix = "arXiv",
    primaryClass = "gr-qc",
    doi = "10.1103/PhysRevD.99.104003",
    journal = "Phys. Rev. D",
    volume = "99",
    number = "10",
    pages = "104003",
    year = "2019"
}

@misc{Skvortsova:2026testfields,
    author = "Skvortsova, Milena",
    title = "{Quasinormal modes and excitation factors of a regular black hole with zero-point length}",
    eprint = "2606.15785",
    archivePrefix = "arXiv",
    primaryClass = "gr-qc",
    year = "2026"
}

@article{Kokkotas:1999bd,
    author = "Kokkotas, Kostas D. and Schmidt, Bernd G.",
    title = "{Quasinormal modes of stars and black holes}",
    eprint = "gr-qc/9909058",
    archivePrefix = "arXiv",
    doi = "10.12942/lrr-1999-2",
    journal = "Living Rev. Rel.",
    volume = "2",
    pages = "2",
    year = "1999"
}

@article{Nollert:1999ji,
    author = "Nollert, Hans-Peter",
    title = "{TOPICAL REVIEW: Quasinormal modes: the characteristic `sound' of black holes and neutron stars}",
    doi = "10.1088/0264-9381/16/12/201",
    journal = "Class. Quant. Grav.",
    volume = "16",
    pages = "R159--R216",
    year = "1999"
}

@article{Bolokhov:2025rng,
    author = "Bolokhov, Sergei V. and Skvortsova, Milena",
    title = "{Review of Analytic Results on Quasinormal Modes of Black Holes}",
    doi = "10.1134/S0202289325700306",
    journal = "Grav. Cosmol.",
    volume = "31",
    number = "4",
    pages = "423--446",
    year = "2025"
}

@article{Konoplya:2022pbc,
    author = "Konoplya, R. A. and Zhidenko, A.",
    title = "{First few overtones probe the event horizon geometry}",
    eprint = "2209.00679",
    archivePrefix = "arXiv",
    primaryClass = "gr-qc",
    doi = "10.1016/j.jheap.2024.10.015",
    journal = "JHEAp",
    volume = "44",
    pages = "419--426",
    year = "2024"
}

@article{Cardoso:2008bp,
    author = "Cardoso, Vitor and Miranda, Alex S. and Berti, Emanuele and Witek, Helvi and Zanchin, Vilson T.",
    title = "{Geodesic stability, Lyapunov exponents and quasinormal modes}",
    eprint = "0812.1806",
    archivePrefix = "arXiv",
    primaryClass = "hep-th",
    doi = "10.1103/PhysRevD.79.064016",
    journal = "Phys. Rev. D",
    volume = "79",
    number = "6",
    pages = "064016",
    year = "2009"
}

@article{Spina:2025wxb,
    author = "Spina, Andrea",
    title = "{Black Holes in Asymptotic Safety: A Review of Solutions and Phenomenology}",
    eprint = "2510.14552",
    archivePrefix = "arXiv",
    primaryClass = "gr-qc",
    doi = "10.53941/ijgtp.2025.100008",
    journal = "Int. J. Grav. Theor. Phys.",
    volume = "1",
    number = "1",
    pages = "8",
    year = "2025"
}

@article{Ayon-Beato:1998hmi,
    author = "Ayon-Beato, Eloy and Garcia, Alberto",
    title = "{Regular black hole in general relativity coupled to nonlinear electrodynamics}",
    eprint = "gr-qc/9911046",
    archivePrefix = "arXiv",
    doi = "10.1103/PhysRevLett.80.5056",
    journal = "Phys. Rev. Lett.",
    volume = "80",
    pages = "5056--5059",
    year = "1998"
}

@article{Bronnikov:2000vy,
    author = "Bronnikov, Kirill A.",
    title = "{Regular magnetic black holes and monopoles from nonlinear electrodynamics}",
    eprint = "gr-qc/0006014",
    archivePrefix = "arXiv",
    doi = "10.1103/PhysRevD.63.044005",
    journal = "Phys. Rev. D",
    volume = "63",
    pages = "044005",
    year = "2001"
}

@article{Konoplya:2025ect,
    author = "Konoplya, R. A. and Zhidenko, A.",
    title = "{Dark matter halo as a source of regular black-hole geometries}",
    eprint = "2511.03066",
    archivePrefix = "arXiv",
    primaryClass = "gr-qc",
    doi = "10.1103/7ptp-9j1t",
    journal = "Phys. Rev. D",
    volume = "113",
    number = "4",
    pages = "043011",
    year = "2026"
}

@article{Nicolini:2005vd,
    author = "Nicolini, Piero and Smailagic, Anais and Spallucci, Euro",
    title = "{Noncommutative geometry inspired Schwarzschild black hole}",
    eprint = "gr-qc/0510112",
    archivePrefix = "arXiv",
    doi = "10.1016/j.physletb.2005.11.004",
    journal = "Phys. Lett. B",
    volume = "632",
    pages = "547--551",
    year = "2006"
}

@article{Bonanno:2025dry,
    author = "Bonanno, Alfio M. and Konoplya, Roman A. and Oglialoro, Giovanni and Spina, Andrea",
    title = "{Regular black holes from proper-time flow in quantum gravity and their quasinormal modes, shadow and Hawking radiation}",
    eprint = "2509.12469",
    archivePrefix = "arXiv",
    primaryClass = "gr-qc",
    doi = "10.1088/1475-7516/2025/12/042",
    journal = "JCAP",
    volume = "12",
    pages = "042",
    year = "2025"
}

@article{Guo:2024jhg,
    author = "Guo, Yang and Xie, Hao and Miao, Yan-Gang",
    title = "{Signal of phase transition hidden in quasinormal modes of regular AdS black holes}",
    eprint = "2402.10406",
    archivePrefix = "arXiv",
    primaryClass = "gr-qc",
    doi = "10.1016/j.physletb.2024.138801",
    journal = "Phys. Lett. B",
    volume = "855",
    pages = "138801",
    year = "2024"
}

@article{3164933,
    author = {L{\"u}tf{\"u}o{\u{g}}lu, Bekir Can and Shermatov, Abubakir and Murodov, Sardor and Javlon, Rayimbaev and Umaraliyev, Maksud},
    title = "{Evaporation of a Regular Black Hole with Gravitational Self-Energy}",
    doi = "10.13140/RG.2.2.34442.22726"
}

@article{Li:2014fka,
    author = "Li, Jin and Lin, Kai and Yang, Nan",
    title = "{Nonlinear electromagnetic quasinormal modes and Hawking radiation of a regular black hole with magnetic charge}",
    eprint = "1409.5988",
    archivePrefix = "arXiv",
    primaryClass = "gr-qc",
    doi = "10.1140/epjc/s10052-015-3347-3",
    journal = "Eur. Phys. J. C",
    volume = "75",
    number = "3",
    pages = "131",
    year = "2015"
}

@article{Bolokhov:2023ruj,
    author = "Bolokhov, S. V.",
    title = "{Long-lived quasinormal modes and oscillatory tails of the Bardeen spacetime}",
    doi = "10.1103/PhysRevD.109.064017",
    journal = "Phys. Rev. D",
    volume = "109",
    number = "6",
    pages = "064017",
    year = "2024"
}

@article{Lopez:2022uie,
    author = "L{\'o}pez, L. A. and Ram{\'\i}rez, Valeria",
    title = "{Quasi-normal modes of a Generic-class of magnetically charged regular black hole: scalar and electromagnetic perturbations}",
    eprint = "2205.10166",
    archivePrefix = "arXiv",
    primaryClass = "gr-qc",
    doi = "10.1140/epjp/s13360-023-03735-6",
    journal = "Eur. Phys. J. Plus",
    volume = "138",
    number = "2",
    pages = "120",
    year = "2023"
}

@article{Panotopoulos:2019qjk,
    author = "Panotopoulos, Grigoris and Rinc{\'o}n, {\'A}ngel",
    title = "{Quasinormal modes of regular black holes with non linear-Electrodynamical sources}",
    eprint = "1904.10847",
    archivePrefix = "arXiv",
    primaryClass = "gr-qc",
    doi = "10.1140/epjp/i2019-12686-x",
    journal = "Eur. Phys. J. Plus",
    volume = "134",
    number = "6",
    pages = "300",
    year = "2019"
}

@article{Skvortsova:2024wly,
    author = "Skvortsova, Milena",
    title = "{Ringing of Extreme Regular Black Holes}",
    eprint = "2405.15807",
    archivePrefix = "arXiv",
    primaryClass = "gr-qc",
    doi = "10.1134/S020228932470018X",
    journal = "Grav. Cosmol.",
    volume = "30",
    number = "3",
    pages = "279--288",
    year = "2024"
}

@article{Flachi:2012nv,
    author = "Flachi, Antonino and Lemos, Jos{\'e} P. S.",
    title = "{Quasinormal modes of regular black holes}",
    eprint = "1211.6212",
    archivePrefix = "arXiv",
    primaryClass = "gr-qc",
    doi = "10.1103/PhysRevD.87.024034",
    journal = "Phys. Rev. D",
    volume = "87",
    number = "2",
    pages = "024034",
    year = "2013"
}

@article{Held:2019xde,
    author = "Held, Aaron and Gold, Roman and Eichhorn, Astrid",
    title = "{Asymptotic safety casts its shadow}",
    eprint = "1904.07133",
    archivePrefix = "arXiv",
    primaryClass = "gr-qc",
    doi = "10.1088/1475-7516/2019/06/029",
    journal = "JCAP",
    volume = "06",
    pages = "029",
    year = "2019"
}

@article{Dubinsky:2026wcv,
    author = "Dubinsky, Alexey",
    title = "{Scattering of a scalar field in the four-dimensional quasi-topological gravity}",
    eprint = "2603.17644",
    archivePrefix = "arXiv",
    primaryClass = "gr-qc",
    doi = "10.53941/ijgtp.2026.100006",
    journal = "Int. J. Grav. Theor. Phys.",
    volume = "2",
    number = "1",
    pages = "6",
    year = "2026"
}

@article{Yang:2021cvh,
    author = "Yang, Yi and Liu, Dong and Xu, Zhaoyi and Xing, Yujia and Wu, Shurui and Long, Zheng-Wen",
    title = "{Echoes of novel black-bounce spacetimes}",
    eprint = "2107.06554",
    archivePrefix = "arXiv",
    primaryClass = "gr-qc",
    doi = "10.1103/PhysRevD.104.104021",
    journal = "Phys. Rev. D",
    volume = "104",
    number = "10",
    pages = "104021",
    year = "2021"
}

@article{Skvortsova:2025cah,
    author = "Skvortsova, Milena",
    title = "{Arbitrarily long-lived quasinormal modes of proper-time flow black holes}",
    eprint = "2509.18061",
    archivePrefix = "arXiv",
    primaryClass = "gr-qc",
    doi = "10.1209/0295-5075/ae5e03",
    journal = "EPL",
    volume = "154",
    number = "4",
    pages = "49001",
    year = "2026"
}

@article{Pedraza:2021hzw,
    author = "Pedraza, Omar and L{\'o}pez, L. A. and Arceo, R. and Cabrera-Munguia, I.",
    title = "{Quasinormal modes of the Hayward black hole surrounded by quintessence: Scalar, electromagnetic and gravitational perturbations}",
    eprint = "2111.06488",
    archivePrefix = "arXiv",
    primaryClass = "gr-qc",
    doi = "10.1142/S0217732322500572",
    journal = "Mod. Phys. Lett. A",
    volume = "37",
    number = "09",
    pages = "2250057",
    year = "2022"
}

@article{Gingrich:2024tuf,
    author = "Gingrich, Douglas M.",
    title = "{Quasinormal modes of a nonsingular spherically symmetric black hole effective model with holonomy corrections}",
    eprint = "2404.04447",
    archivePrefix = "arXiv",
    primaryClass = "gr-qc",
    doi = "10.1103/PhysRevD.110.084045",
    journal = "Phys. Rev. D",
    volume = "110",
    number = "8",
    pages = "084045",
    year = "2024"
}

@article{Huang:2023aet,
    author = "Huang, Bai-Hao and Hu, Han-Wen and Zhao, Liu",
    title = "{Thermodynamics for regular black holes as intermediate thermodynamic states and quasinormal frequencies}",
    eprint = "2311.12286",
    archivePrefix = "arXiv",
    primaryClass = "gr-qc",
    doi = "10.1088/1475-7516/2024/03/053",
    journal = "JCAP",
    volume = "03",
    pages = "053",
    year = "2024"
}

@article{Bolokhov:2025fto,
    author = "Bolokhov, S. V.",
    title = "{Quasinormal ringing of a regular black hole sourced by the Dehnen-type distribution of matter}",
    eprint = "2511.12859",
    archivePrefix = "arXiv",
    primaryClass = "gr-qc",
    doi = "10.1016/j.aop.2026.170416",
    journal = "Annals Phys.",
    volume = "488",
    pages = "170416",
    year = "2026"
}

@article{Skvortsova:2026unq,
    author = "Skvortsova, Milena",
    title = "{Long-lived quasinormal frequencies for regular black hole supported by the Einasto profile in the presence of the magnetic field}",
    eprint = "2603.28415",
    archivePrefix = "arXiv",
    primaryClass = "gr-qc",
    month = "3",
    year = "2026"
}

@article{MahdavianYekta:2019pol,
    author = "Mahdavian Yekta, Davood and Karimabadi, Majid and Alavi, S. A.",
    title = "{Quasinormal modes for non-minimally coupled scalar fields in regular black hole spacetimes: Grey-body factors, area spectrum and shadow radius}",
    eprint = "1912.12017",
    archivePrefix = "arXiv",
    primaryClass = "hep-th",
    doi = "10.1016/j.aop.2021.168603",
    journal = "Annals Phys.",
    volume = "434",
    pages = "168603",
    year = "2021"
}

@article{Lutfuoglu:2026fpx,
    author = {L{\"u}tf{\"u}o{\u{g}}lu, Bekir Can and Rayimbaev, Javlon and Rahmatov, Bekzod and Shayimov, Fayzullo and Davletov, Ikram},
    title = "{Telling tails and quasi-resonances in the vicinity of Dymnikova regular black hole}",
    eprint = "2601.17906",
    archivePrefix = "arXiv",
    primaryClass = "gr-qc",
    doi = "10.1016/j.physletb.2026.140392",
    journal = "Phys. Lett. B",
    volume = "876",
    pages = "140392",
    year = "2026"
}

@article{Konoplya:2023moy,
    author = "Konoplya, R. A. and Zhidenko, A.",
    title = "{Analytic expressions for quasinormal modes and grey-body factors in the eikonal limit and beyond}",
    eprint = "2309.02560",
    archivePrefix = "arXiv",
    primaryClass = "gr-qc",
    doi = "10.1088/1361-6382/ad0a52",
    journal = "Class. Quant. Grav.",
    volume = "40",
    number = "24",
    pages = "245005",
    year = "2023"
}

@article{Matyjasek:2026yiu,
    author = "Matyjasek, Jerzy and Konoplya, Roman A. and Zhidenko, Alexander",
    title = "{An Efficient Higher-Order WKB Code for Quasinormal Modes and Greybody Factors}",
    eprint = "2603.12466",
    archivePrefix = "arXiv",
    primaryClass = "gr-qc",
    doi = "10.53941/ijgtp.2026.100005",
    journal = "Int. J. Grav. Theor. Phys.",
    volume = "2",
    number = "1",
    pages = "5",
    year = "2026"
}

@article{Matyjasek:2019eeu,
    author = "Matyjasek, Jerzy and Telecka, Malgorzata",
    title = "{Quasinormal modes of black holes. II. Pad{\'e} summation of the higher-order WKB terms}",
    eprint = "1908.09389",
    archivePrefix = "arXiv",
    primaryClass = "gr-qc",
    doi = "10.1103/PhysRevD.100.124006",
    journal = "Phys. Rev. D",
    volume = "100",
    number = "12",
    pages = "124006",
    year = "2019"
}

@article{Lutfuoglu:2025pzi,
    author = {L{\"u}tf{\"u}o{\u{g}}lu, Bekir Can and Saka, Erdin{\c{c}} Ula{\c{s}} and Shermatov, Abubakir and Rayimbaev, Javlon and Ibragimov, Inomjon and Muminov, Sokhibjan},
    title = "{Gravitational quasinormal modes of Dymnikova black holes}",
    eprint = "2509.24633",
    archivePrefix = "arXiv",
    primaryClass = "gr-qc",
    doi = "10.1016/j.aop.2026.170360",
    journal = "Annals Phys.",
    volume = "487",
    pages = "170360",
    year = "2026"
}

@article{Lutfuoglu:2025kqp,
    author = {L{\"u}tf{\"u}o{\u{g}}lu, B. C.},
    title = "{Long-lived quasinormal modes, grey-body factors and absorption cross-section of the black hole immersed in the Hernquist galactic halo}",
    eprint = "2510.25969",
    archivePrefix = "arXiv",
    primaryClass = "gr-qc",
    doi = "10.1016/j.physletb.2025.140082",
    journal = "Phys. Lett. B",
    volume = "872",
    pages = "140082",
    year = "2026"
}

@article{Lutfuoglu:2026zel,
    author = {L{\"u}tf{\"u}o{\u{g}}lu, Bekir Can and Rayimbaev, Javlon and Murodov, Sardor and Abdullaev, Mardon and Akhmedov, Munisbek},
    title = "{Ringing Regularity: Gravitational Perturbations and Quasinormal Modes of Einasto-Supported Black Holes}",
    eprint = "2602.20601",
    archivePrefix = "arXiv",
    primaryClass = "gr-qc",
    month = "2",
    year = "2026"
}

@article{Lutfuoglu:2026zxj,
    author = {L{\"u}tf{\"u}o{\u{g}}lu, Bekir Can},
    title = "{Hawking emission from black holes evaporating toward wormholes and the accuracy of the WKB approximation}",
    eprint = "2606.08351",
    archivePrefix = "arXiv",
    primaryClass = "gr-qc",
    doi = "10.1016/j.physletb.2026.140754",
    journal = "Phys. Lett. B",
    volume = "880",
    pages = "140754",
    year = "2026"
}

@article{Bolokhov:2024bke,
    author = "Bolokhov, Sergey",
    title = "{Long lived quasinormal modes in the effective quantum gravity}",
    doi = "10.1140/epjc/s10052-025-14883-7",
    journal = "Eur. Phys. J. C",
    volume = "85",
    number = "10",
    pages = "1166",
    year = "2025"
}

@article{Bolokhov:2024ixe,
    author = "Bolokhov, S. V.",
    title = "{Late time decay of scalar and Dirac fields around an asymptotically de Sitter black hole in the Euler{\textendash}Heisenberg electrodynamics}",
    eprint = "2404.09364",
    archivePrefix = "arXiv",
    primaryClass = "gr-qc",
    doi = "10.1140/epjc/s10052-024-12990-5",
    journal = "Eur. Phys. J. C",
    volume = "84",
    number = "6",
    pages = "634",
    year = "2024"
}

@article{Konoplya:2007zx,
    author = "Konoplya, R. A. and Zhidenko, A.",
    title = "{Decay of a charged scalar and Dirac fields in the Kerr-Newman-de Sitter background}",
    eprint = "0707.1890",
    archivePrefix = "arXiv",
    primaryClass = "hep-th",
    doi = "10.1103/PhysRevD.76.084018",
    journal = "Phys. Rev. D",
    volume = "76",
    number = "8",
    pages = "084018",
    year = "2007",
    note = "[Erratum: Phys.Rev.D 90, 029901 (2014)]"
}

@article{Konoplya:2017tvu,
    author = "Konoplya, Roman A. and Zhidenko, Alexander",
    title = "{Quasinormal modes of massive fermions in Kerr spacetime: Long-lived modes and the fine structure}",
    eprint = "1712.06667",
    archivePrefix = "arXiv",
    primaryClass = "gr-qc",
    doi = "10.1103/PhysRevD.97.084034",
    journal = "Phys. Rev. D",
    volume = "97",
    number = "8",
    pages = "084034",
    year = "2018"
}

@article{Lutfuoglu:2026boa,
    author = {L{\"u}tf{\"u}o{\u{g}}lu, Bekir Can},
    title = "{Scalar, electromagnetic, and Dirac perturbations of regular black holes constituting primordial dark matter}",
    eprint = "2604.24349",
    archivePrefix = "arXiv",
    primaryClass = "gr-qc",
    doi = "10.1088/1475-7516/2026/07/003",
    journal = "JCAP",
    volume = "07",
    pages = "003",
    year = "2026"
}

@article{Lutfuoglu:2026rqe,
    author = {L{\"u}tf{\"u}o{\u{g}}lu, Bekir Can and Rayimbaev, Javlon and Murodov, Sardor and Kurbanov, Jakhongir and Matyoqubov, Muhammad},
    title = "{Bardeen spacetime as quantum corrected black hole: Grey-body factors and quasinormal modes of gravitational perturbations}",
    eprint = "2605.11364",
    archivePrefix = "arXiv",
    primaryClass = "gr-qc",
    month = "5",
    year = "2026"
}

@article{Skvortsova:2026jtx,
    author = "Skvortsova, Milena",
    title = "{Massive scalar quasinormal modes of an asymptotically flat regular black hole supported by a phantom Dirac{\textendash}Born{\textendash}Infeld field}",
    eprint = "2604.25471",
    archivePrefix = "arXiv",
    primaryClass = "gr-qc",
    doi = "10.1016/j.aop.2026.170587",
    journal = "Annals Phys.",
    volume = "492",
    pages = "170587",
    year = "2026"
}

@article{Skvortsova:2026idf,
    author = "Skvortsova, Milena",
    title = "{Quasinormal Spectra of Fields of Various Spin in Asymptotically de Sitter Black Holes within Generalized Proca Theory}",
    eprint = "2605.12113",
    archivePrefix = "arXiv",
    primaryClass = "gr-qc",
    month = "5",
    year = "2026"
}

@article{Kanti:2006ua,
    author = "Kanti, P. and Konoplya, R. A. and Zhidenko, A.",
    title = "{Quasi-Normal Modes of Brane-Localised Standard Model Fields. II. Kerr Black Holes}",
    eprint = "gr-qc/0607048",
    archivePrefix = "arXiv",
    doi = "10.1103/PhysRevD.74.064008",
    journal = "Phys. Rev. D",
    volume = "74",
    pages = "064008",
    year = "2006"
}

@article{Konoplya:2024lir,
    author = "Konoplya, R. A. and Zhidenko, A.",
    title = "{Correspondence between grey-body factors and quasinormal modes}",
    eprint = "2406.11694",
    archivePrefix = "arXiv",
    primaryClass = "gr-qc",
    doi = "10.1088/1475-7516/2024/09/068",
    journal = "JCAP",
    volume = "09",
    pages = "068",
    year = "2024"
}

@article{Malik:2024cgb,
    author = "Malik, Zainab",
    title = "{Correspondence between quasinormal modes and grey-body factors for massive fields in Schwarzschild-de~Sitter spacetime}",
    eprint = "2412.19443",
    archivePrefix = "arXiv",
    primaryClass = "gr-qc",
    doi = "10.1088/1475-7516/2025/04/042",
    journal = "JCAP",
    volume = "04",
    pages = "042",
    year = "2025"
}

@article{Skvortsova:2024msa,
    author = "Skvortsova, Milena",
    title = "{Quantum corrected black holes: testing the correspondence between grey-body factors and quasinormal modes}",
    eprint = "2411.06007",
    archivePrefix = "arXiv",
    primaryClass = "gr-qc",
    doi = "10.1140/epjc/s10052-025-14589-w",
    journal = "Eur. Phys. J. C",
    volume = "85",
    number = "8",
    pages = "854",
    year = "2025"
}

@article{Bolokhov:2024otn,
    author = "Bolokhov, S. V. and Skvortsova, Milena",
    title = "{Correspondence between quasinormal modes and grey-body factors of spherically symmetric traversable wormholes}",
    eprint = "2412.11166",
    archivePrefix = "arXiv",
    primaryClass = "gr-qc",
    doi = "10.1088/1475-7516/2025/04/025",
    journal = "JCAP",
    volume = "04",
    pages = "025",
    year = "2025"
}

@article{Lutfuoglu:2025mqa,
    author = {L{\"u}tf{\"u}o{\u{g}}lu, Bekir Can and Shermatov, Abubakir and Rayimbaev, Javlon and Matyoqubov, Muhammad and Sirajiddin, Otaboyev},
    title = "{Gravitational spectra and wave propagation in regular black holes supported by a Dehnen Halo}",
    eprint = "2511.22366",
    archivePrefix = "arXiv",
    primaryClass = "gr-qc",
    doi = "10.1140/epjc/s10052-025-15234-2",
    journal = "Eur. Phys. J. C",
    volume = "85",
    number = "12",
    pages = "1484",
    year = "2025"
}

@article{Han:2025cal,
    author = "Han, Hyewon and Gwak, Bogeun",
    title = "{Correspondence between quasinormal modes and greybody factors in five-dimensional black holes}",
    eprint = "2508.12989",
    archivePrefix = "arXiv",
    primaryClass = "gr-qc",
    doi = "10.1103/n2ns-drkp",
    journal = "Phys. Rev. D",
    volume = "113",
    number = "6",
    pages = "064058",
    year = "2026"
}

@article{Han:2026fpn,
    author = "Han, Hyewon and Gwak, Bogeun",
    title = "{Correspondence between Quasinormal Modes and Grey-Body Factors of Schwarzschild{\textendash}Tangherlini Black Holes}",
    eprint = "2601.18613",
    archivePrefix = "arXiv",
    primaryClass = "gr-qc",
    doi = "10.1093/ptep/ptag056",
    journal = "PTEP",
    volume = "2026",
    number = "4",
    pages = "043E01",
    year = "2026"
}

@article{Dubinsky:2024vbn,
    author = "Dubinsky, Alexey",
    title = "{Gray-body factors for gravitational and electromagnetic perturbations around Gibbons{\textendash}Maeda{\textendash}Garfinkle{\textendash}Horowitz{\textendash}Strominger black holes}",
    eprint = "2412.00625",
    archivePrefix = "arXiv",
    primaryClass = "gr-qc",
    doi = "10.1142/S0217732325501111",
    journal = "Mod. Phys. Lett. A",
    volume = "40",
    number = "28",
    pages = "2550111",
    year = "2025"
}

@article{Konoplya:2006gq,
    author = "Konoplya, R. A. and Zhidenko, A. and Molina, C.",
    title = "{Late time tails of the massive vector field in a black hole background}",
    eprint = "gr-qc/0602047",
    archivePrefix = "arXiv",
    doi = "10.1103/PhysRevD.75.084004",
    journal = "Phys. Rev. D",
    volume = "75",
    pages = "084004",
    year = "2007"
}

@article{Kodama:2009bf,
    author = "Kodama, Hideo and Konoplya, R. A. and Zhidenko, Alexander",
    title = "{Gravitational stability of simply rotating Myers-Perry black holes: Tensorial perturbations}",
    eprint = "0904.2154",
    archivePrefix = "arXiv",
    primaryClass = "gr-qc",
    doi = "10.1103/PhysRevD.81.044007",
    journal = "Phys. Rev. D",
    volume = "81",
    pages = "044007",
    year = "2010"
}

@article{Konoplya:2007yy,
    author = "Konoplya, R. A. and Fontana, R. D. B.",
    title = "{Quasinormal modes of black holes immersed in a strong magnetic field}",
    eprint = "0707.1156",
    archivePrefix = "arXiv",
    primaryClass = "hep-th",
    doi = "10.1016/j.physletb.2007.10.065",
    journal = "Phys. Lett. B",
    volume = "659",
    pages = "375--379",
    year = "2008"
}

@article{Konoplya:2013sba,
    author = "Konoplya, R. A. and Zhidenko, A.",
    title = "{Instability of D-dimensional extremally charged Reissner-Nordstrom(-de Sitter) black holes: Extrapolation to arbitrary D}",
    eprint = "1309.7667",
    archivePrefix = "arXiv",
    primaryClass = "hep-th",
    doi = "10.1103/PhysRevD.89.024011",
    journal = "Phys. Rev. D",
    volume = "89",
    number = "2",
    pages = "024011",
    year = "2014"
}

@article{Konoplya:2023fmh,
    author = "Konoplya, R. A. and Zhidenko, A.",
    title = "{Asymptotic tails of massive gravitons in light of pulsar timing array observations}",
    eprint = "2307.01110",
    archivePrefix = "arXiv",
    primaryClass = "gr-qc",
    doi = "10.1016/j.physletb.2024.138685",
    journal = "Phys. Lett. B",
    volume = "853",
    pages = "138685",
    year = "2024"
}

@article{Konoplya:2009hv,
    author = "Konoplya, R. A. and Zhidenko, A.",
    title = "{Holographic conductivity of zero temperature superconductors}",
    eprint = "0909.2138",
    archivePrefix = "arXiv",
    primaryClass = "hep-th",
    doi = "10.1016/j.physletb.2010.02.048",
    journal = "Phys. Lett. B",
    volume = "686",
    pages = "199--206",
    year = "2010"
}

@article{Konoplya:2024vuj,
    author = "Konoplya, R. A. and Zhidenko, A.",
    title = "{Correspondence between grey-body factors and quasinormal frequencies for rotating black holes}",
    eprint = "2408.11162",
    archivePrefix = "arXiv",
    primaryClass = "gr-qc",
    doi = "10.1016/j.physletb.2025.139288",
    journal = "Phys. Lett. B",
    volume = "861",
    pages = "139288",
    year = "2025"
}

@article{Dubinsky:2025wns,
    author = "Dubinsky, Alexey",
    title = "{Long-lived Modes and Grey-body Factors of Massive Fields in Quantum-corrected (Hayward) Black Holes}",
    eprint = "2511.00778",
    archivePrefix = "arXiv",
    primaryClass = "gr-qc",
    doi = "10.1007/s10773-026-06274-9",
    journal = "Int. J. Theor. Phys.",
    volume = "65",
    number = "2",
    pages = "45",
    year = "2026"
}

@article{Dubinsky:2025bvf,
    author = "Dubinsky, Alexey",
    title = "{Long-lived quasinormal modes and quasi-resonances around non-minimal Einstein{\textendash}Yang{\textendash}Mills black holes}",
    eprint = "2505.08545",
    archivePrefix = "arXiv",
    primaryClass = "gr-qc",
    doi = "10.1140/epjc/s10052-025-14671-3",
    journal = "Eur. Phys. J. C",
    volume = "85",
    number = "8",
    pages = "924",
    year = "2025"
}

@article{Malik:2024bmp,
    author = "Malik, Zainab",
    title = "{Quasinormal modes of the Mannheim{\textendash}Kazanas black holes}",
    doi = "10.1515/zna-2024-0153",
    journal = "Z. Naturforsch. A",
    volume = "79",
    number = "11",
    pages = "1063--1073",
    year = "2024"
}

@article{Malik:2024tuf,
    author = "Malik, Zainab",
    title = "{Analytical QNMs of fields of various spin in the Hayward spacetime}",
    eprint = "2410.04306",
    archivePrefix = "arXiv",
    primaryClass = "gr-qc",
    reportNumber = "Research Gate Preprint: DOI:10.13140/RG.2.2.32496.06402",
    doi = "10.1209/0295-5075/ad7885",
    journal = "EPL",
    volume = "147",
    number = "6",
    pages = "69001",
    year = "2024"
}

@article{Bolokhov:2026eqf,
    author = "Bolokhov, S. V.",
    title = "{Quasinormal Modes and Grey-Body Factors of Scalar, Electromagnetic and Dirac Fields Around Einasto-Supported Regular Black Holes}",
    eprint = "2603.22310",
    archivePrefix = "arXiv",
    primaryClass = "gr-qc",
    month = "3",
    year = "2026"
}

@article{Bolokhov:2026uol,
    author = "Bolokhov, S. V.",
    title = "{Massive Scalar Quasinormal Modes, Greybody Factors, and Absorption Cross Section of a Parity-Symmetric Beyond-Horndeski Black Hole}",
    eprint = "2605.11013",
    archivePrefix = "arXiv",
    primaryClass = "gr-qc",
    month = "5",
    year = "2026"
}

@article{Bolokhov:2026dfg,
    author = "Bolokhov, S. V.",
    title = "{Long-lived quasinormal modes of Asymptotically de Sitter Black Holes in Generalized Proca Theory}",
    eprint = "2605.21533",
    archivePrefix = "arXiv",
    primaryClass = "gr-qc",
    month = "5",
    year = "2026"
}

@article{Malik:2026jzl,
    author = "Malik, Zainab",
    title = {{Long-lived massive scalar modes, grey-body factors, and absorption cross sections of the Reissner--Nordstr{\"o}m-like brane-world black hole}},
    eprint = "2605.03659",
    archivePrefix = "arXiv",
    primaryClass = "gr-qc",
    month = "5",
    year = "2026"
}

@article{Malik:2023bxc,
    author = "Malik, Zainab",
    title = "{Quasinormal Modes of the Bumblebee Black Holes with a Global Monopole}",
    eprint = "2308.10412",
    archivePrefix = "arXiv",
    primaryClass = "gr-qc",
    doi = "10.1007/s10773-024-05737-1",
    journal = "Int. J. Theor. Phys.",
    volume = "63",
    number = "8",
    pages = "199",
    year = "2024"
}

@article{Malik:2024iky,
    author = "Malik, Zainab",
    title = "{Analytic Expressions for Quasinormal Modes in Einstein{\textendash}Aether Theory}",
    doi = "10.1134/S0202289325700598",
    journal = "Grav. Cosmol.",
    volume = "32",
    number = "1",
    pages = "122--134",
    year = "2026"
}

@article{Dubinsky:2024gwo,
    author = "Dubinsky, Alexey",
    title = "{Overtones of black holes via time-domain integration}",
    eprint = "2404.18004",
    archivePrefix = "arXiv",
    primaryClass = "gr-qc",
    doi = "10.1142/S0217732324501086",
    journal = "Mod. Phys. Lett. A",
    volume = "39",
    number = "21n22",
    pages = "2450108",
    year = "2024"
}

@article{Dubinsky:2024jqi,
    author = "Dubinsky, Alexey",
    title = "{Telling late-time tails for a massive scalar field in the background of brane-localized black holes}",
    eprint = "2403.01883",
    archivePrefix = "arXiv",
    primaryClass = "gr-qc",
    doi = "10.1209/0295-5075/ad51a3",
    journal = "EPL",
    volume = "147",
    number = "1",
    pages = "19003",
    year = "2024"
}

@article{Dubinsky:2024hmn,
    author = "Dubinsky, Alexey and Zinhailo, Antonina",
    title = "{Asymptotic decay and quasinormal frequencies of scalar and Dirac fields around dilaton-de Sitter black holes}",
    eprint = "2404.01834",
    archivePrefix = "arXiv",
    primaryClass = "gr-qc",
    doi = "10.1140/epjc/s10052-024-13206-6",
    journal = "Eur. Phys. J. C",
    volume = "84",
    number = "8",
    pages = "847",
    year = "2024"
}

@article{Dubinsky:2024rvf,
    author = "Dubinsky, Alexey",
    title = "{Analytic expressions for quasinormal modes of the general parametrized spherically symmetric black holes and the Hod's proposal}",
    eprint = "2409.16569",
    archivePrefix = "arXiv",
    primaryClass = "gr-qc",
    doi = "10.1016/j.physletb.2025.139251",
    journal = "Phys. Lett. B",
    volume = "861",
    pages = "139251",
    year = "2025"
}

@article{Dubinsky:2025nxv,
    author = "Dubinsky, Alexey",
    title = "{Gravitational perturbations of Dymnikova black holes: Grey-body factors and absorption cross-sections}",
    eprint = "2509.11017",
    archivePrefix = "arXiv",
    primaryClass = "gr-qc",
    doi = "10.1016/j.aop.2025.170299",
    journal = "Annals Phys.",
    volume = "485",
    pages = "170299",
    year = "2026"
}

@article{Dubinsky:2025ypj,
    author = "Dubinsky, Alexey",
    title = "{Scattering and absorption of Standard Model fields by brane-localized Schwarzschild{\textendash}de Sitter black holes}",
    eprint = "2510.11643",
    archivePrefix = "arXiv",
    primaryClass = "gr-qc",
    doi = "10.1142/s0218271826500094",
    journal = "Int. J. Mod. Phys. D",
    volume = "35",
    number = "07",
    pages = "2650009",
    year = "2026"
}

@article{Konoplya:2019ppy,
    author = "Konoplya, R. A. and Zinhailo, A. F.",
    title = "{Hawking radiation of non-Schwarzschild black holes in higher derivative gravity: a crucial role of grey-body factors}",
    eprint = "1904.05341",
    archivePrefix = "arXiv",
    primaryClass = "gr-qc",
    doi = "10.1103/PhysRevD.99.104060",
    journal = "Phys. Rev. D",
    volume = "99",
    number = "10",
    pages = "104060",
    year = "2019"
}

@article{Skvortsova:2023zca,
    author = "Skvortsova, Milena",
    title = "{Stability of Asymptotically Flat $\mathbf{(2+1)}$-Dimensional Black Holes with Gauss{\textendash}Bonnet Corrections}",
    eprint = "2311.02729",
    archivePrefix = "arXiv",
    primaryClass = "gr-qc",
    doi = "10.1134/S0202289324010110",
    journal = "Grav. Cosmol.",
    volume = "30",
    number = "1",
    pages = "68--70",
    year = "2024"
}

@article{Konoplya:2025hgp,
    author = "Konoplya, R. A. and Stashko, O. S.",
    title = "{Transition from regular black holes to wormholes in covariant effective quantum gravity: Scattering, quasinormal modes, and Hawking radiation}",
    eprint = "2502.05689",
    archivePrefix = "arXiv",
    primaryClass = "gr-qc",
    doi = "10.1103/PhysRevD.111.084031",
    journal = "Phys. Rev. D",
    volume = "111",
    number = "8",
    pages = "084031",
    year = "2025"
}

@article{Dymnikova:1992ux,
    author = "Dymnikova, I.",
    title = "{Vacuum nonsingular black hole}",
    doi = "10.1007/BF00760226",
    journal = "Gen. Rel. Grav.",
    volume = "24",
    pages = "235--242",
    year = "1992"
}

@article{Konoplya:2024kih,
    author = "Konoplya, R. A. and Zhidenko, A.",
    title = "{Dymnikova black hole from an infinite tower of higher-curvature corrections}",
    eprint = "2404.09063",
    archivePrefix = "arXiv",
    primaryClass = "gr-qc",
    doi = "10.1016/j.physletb.2024.138945",
    journal = "Phys. Lett. B",
    volume = "856",
    pages = "138945",
    year = "2024"
}

@article{Arbey:2021jif,
    author = "Arbey, Alexandre and Auffinger, J{\'e}r{\'e}my and Geiller, Marc and Livine, Etera R. and Sartini, Francesco",
    title = "{Hawking radiation by spherically-symmetric static black holes for all spins: Teukolsky equations and potentials}",
    eprint = "2101.02951",
    archivePrefix = "arXiv",
    primaryClass = "gr-qc",
    reportNumber = "CERN-TH-2021-007",
    doi = "10.1103/PhysRevD.103.104010",
    journal = "Phys. Rev. D",
    volume = "103",
    number = "10",
    pages = "104010",
    year = "2021"
}

@article{Arbey:2021yke,
    author = "Arbey, Alexandre and Auffinger, J{\'e}r{\'e}my and Geiller, Marc and Livine, Etera R. and Sartini, Francesco",
    title = "{Hawking radiation by spherically-symmetric static black holes for all spins. II. Numerical emission rates, analytical limits, and new constraints}",
    eprint = "2107.03293",
    archivePrefix = "arXiv",
    primaryClass = "gr-qc",
    reportNumber = "CERN-TH-2021-105",
    doi = "10.1103/PhysRevD.104.084016",
    journal = "Phys. Rev. D",
    volume = "104",
    number = "8",
    pages = "084016",
    year = "2021"
}

@article{Arbey:2026koc,
    author = "Arbey, Alexandre and Calz{\`a}, Marco and Malacher, L{\'e}a and Pedrotti, Davide and Perez-Gonzalez, Yuber F.",
    title = "{BlackHawk v3.0: Hawking radiation from regular black holes}",
    eprint = "2606.06355",
    archivePrefix = "arXiv",
    primaryClass = "gr-qc",
    doi = "10.1016/j.dark.2026.102407",
    journal = "Phys. Dark Univ.",
    volume = "53",
    pages = "102407",
    year = "2026"
}

@article{Vagnozzi:2022moj,
    author = "Vagnozzi, Sunny and others",
    title = "{Horizon-scale tests of gravity theories and fundamental physics from the Event Horizon Telescope image of Sagittarius A}",
    eprint = "2205.07787",
    archivePrefix = "arXiv",
    primaryClass = "gr-qc",
    reportNumber = "UCI-HEP-TR-2022-07",
    doi = "10.1088/1361-6382/acd97b",
    journal = "Class. Quant. Grav.",
    volume = "40",
    number = "16",
    pages = "165007",
    year = "2023"
}

@article{Calza:2024xdh,
    author = "Calz{\`a}, Marco and Pedrotti, Davide and Vagnozzi, Sunny",
    title = "{Primordial regular black holes as all the dark matter. II. Non-time-radial-symmetric and loop quantum gravity-inspired metrics}",
    eprint = "2409.02807",
    archivePrefix = "arXiv",
    primaryClass = "gr-qc",
    doi = "10.1103/PhysRevD.111.024010",
    journal = "Phys. Rev. D",
    volume = "111",
    number = "2",
    pages = "024010",
    year = "2025"
}

@article{Pedrotti:2024znu,
    author = "Pedrotti, Davide and Vagnozzi, Sunny",
    title = "{Quasinormal modes-shadow correspondence for rotating regular black holes}",
    eprint = "2404.07589",
    archivePrefix = "arXiv",
    primaryClass = "gr-qc",
    doi = "10.1103/PhysRevD.110.084075",
    journal = "Phys. Rev. D",
    volume = "110",
    number = "8",
    pages = "084075",
    year = "2024"
}

@article{Bronnikov:2024izh,
    author = "Bronnikov, Kirill A.",
    title = "{Regular black holes as an alternative to black bounce}",
    eprint = "2404.14816",
    archivePrefix = "arXiv",
    primaryClass = "gr-qc",
    doi = "10.1103/PhysRevD.110.024021",
    journal = "Phys. Rev. D",
    volume = "110",
    number = "2",
    pages = "024021",
    year = "2024"
}

@article{Bolokhov:2024sdy,
    author = "Bolokhov, S. V. and Bronnikov, K. A. and Skvortsova, M. V.",
    title = "{A Regular Center Instead of a Black Bounce}",
    eprint = "2405.09124",
    archivePrefix = "arXiv",
    primaryClass = "gr-qc",
    doi = "10.1134/S0202289324700178",
    journal = "Grav. Cosmol.",
    volume = "30",
    number = "3",
    pages = "265--278",
    year = "2024"
}

@article{Price:1972pw,
    author = "Price, Richard H.",
    title = "{Nonspherical perturbations of relativistic gravitational collapse. I. Scalar and gravitational perturbations}",
    doi = "10.1103/PhysRevD.5.2419",
    journal = "Phys. Rev. D",
    volume = "5",
    pages = "2419--2438",
    year = "1972"
}

@article{Bouhmadi-Lopez:2020oia,
    author = "Bouhmadi-L{\'o}pez, Mariam and Brahma, Suddhasattwa and Chen, Che-Yu and Chen, Pisin and Yeom, Dong-han",
    title = "{A consistent model of non-singular Schwarzschild black hole in loop quantum gravity and its quasinormal modes}",
    eprint = "2004.13061",
    archivePrefix = "arXiv",
    primaryClass = "gr-qc",
    doi = "10.1088/1475-7516/2020/07/066",
    journal = "JCAP",
    volume = "07",
    pages = "066",
    year = "2020"
}

@article{Bolokhov:2025egl,
    author = "Bolokhov, S. V. and Skvortsova, Milena",
    title = "{Gravitational quasinormal modes of the Hayward spacetime}",
    eprint = "2508.19989",
    archivePrefix = "arXiv",
    primaryClass = "gr-qc",
    doi = "10.1140/epjc/s10052-026-15624-0",
    journal = "Eur. Phys. J. C",
    volume = "86",
    number = "4",
    pages = "374",
    year = "2026"
}

@article{Lutfuoglu:2026uzy,
    author = {L{\"u}tf{\"u}o{\u{g}}lu, Bekir Can},
    title = "{Quasinormal Modes of a Massive Scalar Field in 4D Einstein--Gauss--Bonnet Black Hole Spacetimes}",
    eprint = "2603.24424",
    archivePrefix = "arXiv",
    primaryClass = "gr-qc",
    month = "3",
    year = "2026"
}

@article{Ohashi:2004wr,
    author = "Ohashi, Akira and Sakagami, Masa-aki",
    title = "{Massive quasi-normal mode}",
    eprint = "gr-qc/0407009",
    archivePrefix = "arXiv",
    doi = "10.1088/0264-9381/21/16/010",
    journal = "Class. Quant. Grav.",
    volume = "21",
    pages = "3973--3984",
    year = "2004"
}

@article{Koyama:2001qw,
    author = "Koyama, Hiroko and Tomimatsu, Akira",
    title = "{Slowly decaying tails of massive scalar fields in spherically symmetric space-times}",
    eprint = "gr-qc/0112075",
    archivePrefix = "arXiv",
    doi = "10.1103/PhysRevD.65.084031",
    journal = "Phys. Rev. D",
    volume = "65",
    pages = "084031",
    year = "2002"
}

@article{Moderski:2001tk,
    author = "Moderski, Rafal and Rogatko, Marek",
    title = "{Late time evolution of a selfinteracting scalar field in the space-time of dilaton black hole}",
    eprint = "gr-qc/0105056",
    archivePrefix = "arXiv",
    doi = "10.1103/PhysRevD.64.044024",
    journal = "Phys. Rev. D",
    volume = "64",
    pages = "044024",
    year = "2001"
}

@article{Rogatko:2007zz,
    author = "Rogatko, Marek and Szyplowska, Agnieszka",
    title = "{Decay of massive scalar hair on brane black holes}",
    doi = "10.1103/PhysRevD.76.044010",
    journal = "Phys. Rev. D",
    volume = "76",
    pages = "044010",
    year = "2007"
}

@article{Konoplya:2002wt,
    author = "Konoplya, R. A.",
    title = "{Massive charged scalar field in a Reissner-Nordstrom black hole background: Quasinormal ringing}",
    eprint = "gr-qc/0210105",
    archivePrefix = "arXiv",
    doi = "10.1016/S0370-2693(02)02974-X",
    journal = "Phys. Lett. B",
    volume = "550",
    pages = "117--120",
    year = "2002"
}

@article{Lutfuoglu:2026gis,
    author = {L{\"u}tf{\"u}o{\u{g}}lu, Bekir Can},
    title = "{Long-lived quasinormal modes, shadows and particle motion in four-dimensional quasi-topological gravity}",
    eprint = "2603.10844",
    archivePrefix = "arXiv",
    primaryClass = "gr-qc",
    doi = "10.1140/epjc/s10052-026-15807-9",
    journal = "Eur. Phys. J. C",
    volume = "86",
    number = "5",
    pages = "515",
    year = "2026"
}

@article{Bonanno:2000ep,
    author = "Bonanno, Alfio and Reuter, Martin",
    title = "{Renormalization group improved black hole space-times}",
    eprint = "hep-th/0002196",
    archivePrefix = "arXiv",
    reportNumber = "INFN-CT-03-00, MZ-TH-00-04",
    doi = "10.1103/PhysRevD.62.043008",
    journal = "Phys. Rev. D",
    volume = "62",
    pages = "043008",
    year = "2000"
}

@article{Hatsuda:2019eoj,
    author = "Hatsuda, Yasuyuki",
    title = "{Quasinormal modes of black holes and Borel summation}",
    eprint = "1906.07232",
    archivePrefix = "arXiv",
    primaryClass = "gr-qc",
    reportNumber = "RUP-19-18",
    doi = "10.1103/PhysRevD.101.024008",
    journal = "Phys. Rev. D",
    volume = "101",
    number = "2",
    pages = "024008",
    year = "2020"
}
\end{document}